# On the probability of invalidating a causal inference due to limited external validity

Tenglong Li

Michigan State University

## Abstract

External validity is often questionable in empirical research, especially in randomized experiments due to the trade-off between internal validity and external validity. To quantify the robustness of external validity, one must first conceptualize the gap between a sample that is fully representative of the target population (i.e., the ideal sample) and the observed sample. Drawing on Frank & Min (2007) and Frank et al. (2013), I define such gap as the unobserved sample and intend to quantify its relationship with the null hypothesis statistical testing (NHST) in this study. The probability of invalidating a causal inference due to limited external validity, i.e., the PEV, is the probability of failing to reject the null hypothesis based on the ideal sample provided the null hypothesis has been rejected based on the observed sample. This study illustrates the guideline and the procedure of evaluating external validity with the PEV through an empirical example (i.e., Borman et al. (2008)). Specifically, one would be able to locate the threshold of the unobserved sample statistic that would make the PEV higher than a desired value and use this information to characterize the unobserved sample that would render external validity of the research in question less robust. The PEV is shown to be linked to statistical power when the NHST is thought to be based on the ideal sample.

*Keywords:* Hypothesis Testing, Bayesian, Frequentist, Sampling, Regression.

## Introduction

Randomized experiments have long been the benchmark for making causal inferences (Fisher, 1937; Rubin, 2007, 2008; Shadish et al, 2002, 2011; Cook et al. 2010; Morgan & Rubin, 2012; What Works Clearinghouse, 2014; Imbens & Rubin, 2015; Thomas, 2016). However, a randomized experiment usually suffers from a nonrepresentative sample or a convenient sample because it is unethical to randomly sample a human subject and then randomly assign that person to a treatment (Cronbach, 1975; Reichardt & Gollob, 1999; Cook, 2002, 2003, 2007; Schneider et al. 2007; Imai et al. 2008; Stuart et al. 2011; Olsen et al. 2013; Tipton, 2014; O'Muircheartaigh & Hedges, 2014). One prominent example is Borman et al. (2008)., which examined the causal effect of Open Court Reading (OCR) curriculum (National Reading Panel, 2000) by randomly assigning classrooms in each sampled school and each grade to the treatment (OCR) group and the control (non-OCR) group. Their sampled schools were randomly drawn from schools which volunteered for this study, and they may not have represented classrooms in non-volunteer schools (Frank et al., 2013). Cook (2002, 2003, 2007) has formalized such risk as the potential trade-off between internal validity and external validity, i.e., one increase internal validity at the cost of reducing external validity.

The loss of external validity could limit replicability (Garcia & Wantchekon, 2010; Hedges, 2013; Avellar et al., 2017). Considering the difficulty and cost of implementing a randomized experiment, one naturally would (and should) expect the payoff worth the investment, which suggests a significant randomized experiment should be able to generalize widely (Cook et al., 2010; Orr, 2015; Tipton & Peck, 2017; Stuart & Rhodes, 2017). To respond, assessing the evidence regarding external validity has been recommended as an indispensable part of causal research and literature on providing such assessment tools has emerged (e.g., Stuart et al., 2011; O'Muircheartaigh & Hedges, 2014; Tipton, 2014).

Bearing the same goal of indexing the external validity of randomized experiment in mind, this paper approaches it differently. In particular, this article addresses concerns about external validity by considering how a sample would have to change to invalidate an inference. To do so, I consider the target population of interest to consist of an observable population from which the observed sample was taken, and an unobservable population from which no data were sampled but to which one would like to generalize. For example, in order to generalize the inference of Borman et al. (2008) to all American schools, one needs to conceptualize the unobserved sample of classrooms which would be potentially randomly drawn from the non-volunteer schools. Naturally, one would wonder, whether Borman et al. could have found a significant positive effect of OCR again had such unobserved sample became available to them. To answer this question, this article proposes the probability of invalidating a causal inference due to limited external validity (henceforth we denote the probability of invalidating a causal inference as the PEV) as a probabilistic measure of external validity, which is novel for the current literature (Frank and Min 2007; Frank et al., 2013). Furthermore, readers may be interested in learning how the unobserved sample is related to the hypothesis testing that is founded on both the observed and the unobserved samples. Therefore, this article purposes establishing the relationship between the PEV and the unobserved sample statistics that are sufficient for null hypothesis statistical testing (NHST).

This article is organized as follows: The research setting will be firstly outlined and then the definitions of the unobserved and ideal samples are formalized. Drawing on these definitions, the PEV can be formally defined and the relationships between the PEV and the unobserved sample for the simple group-mean-difference and the regression estimators of an average treatment effect are discussed. To illustrate the analytical procedure pertaining to the PEV, the robustness of Borman et al.'s inference (2008) is evaluated in a step-by-step fashion.

## Research setting

### Research design

This article assumes a randomized experiment with a non-representative sample from the target population. Moreover, in this randomized experiment participants are randomly assigned to either the treatment group or the control group. Therefore, there are two groups in total in this prototypical randomized experiment.

### Notation

Notation rule: throughout this article I use a superscript to signal the sample that a statistic pertains to and a subscript to signal the group or the variable(s) that a statistic pertains to. A superscript could be "ob", "un", or "id", which represents the observed sample, the unobserved sample or the ideal sample respectively. A subscript "t" denotes the treatment group and a subscript "c" denotes the control group. A subscript can also be a combination of symbols of variables when it is used to denote variances, covariances or correlations.

Outcomes, sample sizes and mean outcomes: Outcomes are denoted by Y and sample sizes are denoted by n with appropriate superscript and subscript. Mean outcomes are symbolized by $\bar{Y}$ with appropriate superscript and subscript.

Data matrices: **Y** is a vector of outcomes. **Z** is a matrix of covariates. **W** is a vector of treatment status indictor (W=1: Treatment; W=0: Control). An observation should be a vector consisting of the outcome Y, the treatment indicator W and covariates $Z_1$, $Z_2$, …, $Z_p$. **D** denotes a data matrix whose rows are either outcome values (for the simple estimator) or observations (for the regression estimator).

Variances, covariances and correlations: σ and r are used to denote respectively a covariance and a correlation between two variables, with some exceptions which will covered later. **S** and **R** are used to denote a variance-covariance matrix and a correlation matrix respectively. Likewise, all variance, covariance and correlation symbols take appropriate superscripts and subscripts.

**Estimation**

Although there are other forms of causal effect such as the treatment effect for treated or the treatment effect for control, this article only concerns the estimate of an average treatment effect. In addition, this article discusses two different estimators of average treatment effect, i.e., the simple group-mean-difference estimator (referred to as the simple estimator henceforth) and the regression estimator. Specifically, the simple estimator is computed as the difference between the mean outcome for the treatment group and the mean outcome for the control group. The regression estimator is just the estimated regression coefficient of W in the regression of Y on W and $Z_1$, $Z_2$, …, $Z_p$., i.e., the regression estimator is $\hat{\beta}_W$ for the regression model

$$Y = \beta_0 + \beta_W W + \beta_1 Z_1 + \cdots + \beta_p Z_p + \varepsilon .$$

**Hypothesis testing**

To quantify the robustness of an inference, it is necessary to assume a null hypothesis $H_0$: $\delta = \delta_0$ (for the most cases $\delta_0 = 0$; $\delta$ denotes the true average treatment effect) which has been tested and rejected in the observed sample. Moreover, this article assume researchers have subsequently inferred causal effect for their target population which may include populations other than those sampled from.

## The unobserved sample in a thought experiment

To formalize the discussion about the inference of an effect from a randomized experiment based on a sample that is not representative of a full target population, I have the following definitions:

**Definition 1**: **The unobserved part of the target population** refers to the part of the target population from which the observed sample could not be drawn. Conversely, **the observed part of the target population** refers to the part of the target population from which the observed sample was drawn.

**Example:** The unobserved part of the target population of Borman et al. (2008) would be the collection of all classrooms in the non-volunteer schools. The observed part of the target population of Borman et al. (2008) would be the collection of all classrooms in the volunteer schools.

**Definition 2**: **The unobserved sample** refers to an imaginary random sample which is drawn from the unobserved part of the target population. **The unobserved treated sample** refers to the subjects that were randomly assigned to the treatment group in this imaginary random sample. **The unobserved control sample** refers to the subjects that were randomly assigned to the control group in this imaginary random sample.

**Example:** According to definition 2, the unobserved sample of Borman et al. (2008) is an imaginary sample of classrooms which were randomly drawn from non-volunteer schools and subsequently randomly assigned to the Open Court Reading (OCR) group or the control group. This unobserved sample comprises the unobserved treated sample which consists of the classrooms that were randomly assigned to the OCR group and the unobserved control sample which consists of the classrooms that were randomly assigned to the control group.

The above concepts are illustrated by figures 1. In figure 1, the target population of Borman et al. is partitioned into the collection of all classrooms in the volunteer schools (represented by the left panel) and the collection of all classrooms in the non-volunteer schools (represented by the right panel). The conceptualization of unobserved sample is represented by the two arrows which start from the observed sample (the rectangle on the left with blue shaded circles) and end at the unobserved sample (the rectangle on the right with unshaded circles). To conceptualize the unobserved treated sample as well as the unobserved control sample, we can imagine some classrooms were randomly drawn from the non-volunteer schools and subsequently randomly assigned to the OCR group (represented by the upper panel with the label "O") or the control group (represented by the lower panel with the label "C"), following the same sampling and random assignment procedure in Borman et al. (2008).

(Insert Figure 1 here)

How is the unobserved sample related to the bias associated with a non-representative sample in a randomized experiment? Frank & Min (2007) and Frank et al. (2013) proposed that such bias is induced by the gap between the observed sample and the ideal sample one would use for inference, and as we learned from the earlier definitions this gap is represented by the unobserved sample. The "ideal sample" is defined as follows:

**Definition 3**: **The ideal sample** refers to the combination of the observed sample and the unobserved sample. **The ideal treated sample** refers to the combination of the observed treated sample and the unobserved treated sample. **The ideal control sample** refers to the combination of the observed control sample and the unobserved control sample.

Frank & Min (2007) noted that one should think about the ideal sample when his inference is based on a randomized experiment whose observed sample is non-representative of the target population. The spirit of their approach is to reconsider the inference as if the ideal sample were available. Following this spirit, this article intends to achieve two goals: First, this article seeks to build a framework where the inference based on the ideal sample can be *distributionally* equated with the inference based on the observed sample as well as the conceptualization of the unobserved sample. Second, this article seeks to uncover the relationship between the PEV and the unobserved sample statistics, which would allow researchers not only identify thresholds of the unobserved sample statistics that makes the PEV below a desired value (so that external validity is deemed as robust) but also study the overall pattern of how the PEV changes in a broad range of the unobserved samples.

**The probability of invalidating a causal inference due to limited external validity**

Frank et al. (2013) formalized the decision rule for determining when a causal inference is invalid due to limited external validity as follows:

When the null hypothesis has been rejected and a significant positive effect has been inferred, the decision rule is:

$$\hat{\delta} > \delta^{\#} > \delta \tag{1}$$

When the null hypothesis has been rejected and a significant negative effect has been inferred, the decision rule is:

$$\hat{\delta} < \delta^{\#} < \delta \tag{2}$$

Here $\hat{\delta}$ is the estimate of average treatment effect based on the observed sample; $\delta^{\#}$ is the threshold of making an inference (e.g., the threshold defining statistical significance); $\delta$ is a parameter which represents the true average treatment effect. Since the observed sample is fixed in this context, we can treat $\hat{\delta}$ as fixed as well and thus the decision rule can be simplified as follows:

Given a significant positive effect has been concluded, i.e., $\hat{\delta} > \delta^{\#}$ , an inference would be invalidated if:

$$\delta < \delta^{\#} \tag{3}$$

Given a significant negative effect has been inferred, i.e., $\hat{\delta} < \delta^{\#}$, an inference would be invalidated if:

$$\delta > \delta^{\#} \tag{4}$$

Drawing on the decision rule in (3) or (4), the probability of invalidating a causal inference due to limited external validity (PEV) is defined as follows, for the ideal sample:

The PEV is defined as follows based on (3):

$$P(\delta < \delta^{\#} \mid \mathbf{D^{id}}) \tag{5}$$

The PEV is defined as follows based on (4):

$$P(\delta > \delta^{\#} \mid \mathbf{D^{id}}) \tag{6}$$

It is noteworthy that the PEV in (5) and (6) could be alternatively written as $P(\delta < \delta^{\#} \mid \hat{\delta} > \delta^{\#}, \mathbf{D^{id}})$ and $P(\delta > \delta^{\#} \mid \hat{\delta} < \delta^{\#}, \mathbf{D^{id}})$ respectively, since we assume the null hypothesis has been rejected in the observed sample. However, those alternative expressions of the PEV are a bit redundant as the ideal sample $\mathbf{D^{id}}$ incorporates the fixed observed sample and thus the information that the null hypothesis has been rejected. The PEV essentially is the probability of failing to reject the null hypothesis using the ideal sample provided the same null hypothesis has been rejected in the observed sample.

To generate distributional statements, the PEV can be computed over the distribution of $\delta$ conditional on the ideal sample $\mathbf{D^{id}}$, which is often derivable from the sampling distribution of the estimator of $\delta$ based on the ideal sample. One issue remains is that the ideal sample is not accessible to us (let alone the distribution of $\delta$ based on it). To address this issue, this article offers two approaches: the frequentist approach and the Bayesian approach. I will demonstrate that these two approaches converge to the same distribution of $\delta$ when appropriate assumptions are made.

## The relationship between the PEV and the unobserved sample for the simple estimator

### The frequentist approach

**Theorem 1:** Assuming the treated outcome and the control outcome are independent and the variances of those two outcomes are given as $\sigma_t^2$ and $\sigma_c^2$ respectively, the distribution of $\delta$ based on the ideal sample would be:

$$\delta \mid \mathbf{D^{id}} \sim N(\theta_t - \theta_c, \phi_t + \phi_c) \tag{7}$$

Where:

$$\theta_t = \frac{n_t^{un}}{n_t^{ob} + n_t^{un}} \bar{Y}_t^{un} + \frac{n_t^{ob}}{n_t^{ob} + n_t^{un}} \bar{Y}_t^{ob}$$

$$\theta_c = \frac{n_c^{un}}{n_c^{ob} + n_c^{un}} \bar{Y}_c^{un} + \frac{n_c^{ob}}{n_c^{ob} + n_c^{un}} \bar{Y}_c^{ob}$$

$$\phi_t = \frac{\sigma_t^2}{n_t^{ob} + n_t^{un}} \quad (8)$$

$$\phi_c = \frac{\sigma_c^2}{n_c^{ob} + n_c^{un}}$$

given values of $\bar{Y}_t^{un}$ (the unobserved treated sample mean), $\bar{Y}_c^{un}$ (the unobserved control sample mean), $n_t^{un}$ (the unobserved treated sample size) and $n_c^{un}$ (the unobserved control sample size).

(Proof in Appendix).

The expressions in (8) can be simplified if the relative size of the observed part of the target population is given as $\pi_R$. For example, $\pi_R$ would be the proportion of volunteer schools (and arguably schools which are similar to volunteer schools) in Borman et al. (2008) in the population of all U.S. schools. In this case, $n_t^{un} = \left(\frac{1-\pi_R}{\pi_R}\right) n_t^{ob}$ and $n_c^{un} = \left(\frac{1-\pi_R}{\pi_R}\right) n_c^{ob}$, which implies that the proportion of treated subjects in the unobserved sample is expected to be equal to the proportion of treated subjects in the observed sample. Consequently, (8) is rewritten as follows:

$$\begin{aligned}
\theta_t &= (1-\pi_R)\bar{Y}_t^{un} + \pi_R \bar{Y}_t^{ob} \\
\phi_t &= \pi_R \frac{\sigma_t^2}{n_t^{ob}} \\
\theta_c &= (1-\pi_R)\bar{Y}_c^{un} + \pi_R \bar{Y}_c^{ob} \\
\phi_c &= \pi_R \frac{\sigma_c^2}{n_c^{ob}}
\end{aligned} \tag{9}$$

Theorem 1 provides the distribution of $\delta$ conditional on the ideal sample. The observed sample is fixed whereas the unobserved sample could be changed through the manipulation of the unobserved sample statistics in thought experiments. To uncover the relationship between the PEV and the unobserved sample statistics, theorem 2 is presented next:

**Theorem 2:** Drawing on the distribution of $\delta$ in (7) and the expressions in (9), the probit of PEV is a function of α (defined as $\bar{Y}_t^{un} / \bar{Y}_c^{un}$ ), $\bar{Y}_c^{un}$ and $\pi_R$ . Specifically, when a significant positive effect has been concluded, we have:

$$probit(PEV) = \frac{1}{\sqrt{\frac{\sigma_t^2}{n_t^{ob}} + \frac{\sigma_c^2}{n_c^{ob}}}} \left[ \bar{Y}_c^{un} \cdot \alpha \pi_R^{\frac{1}{2}} - \bar{Y}_c^{un} \cdot \alpha \pi_R^{-\frac{1}{2}} - (\bar{Y}_t^{ob} - \bar{Y}_c^{ob} + \bar{Y}_c^{un}) \cdot \pi_R^{\frac{1}{2}} + (\bar{Y}_c^{un} + \delta^{\#}) \cdot \pi_R^{-\frac{1}{2}} \right]$$

(10)

When a significant negative effect has been concluded, we have:

$$probit(PEV) = \frac{1}{\sqrt{\frac{\sigma_t^2}{n_t^{ob}} + \frac{\sigma_c^2}{n_c^{ob}}}} \left[ \bar{Y}_c^{un} \cdot \alpha\pi_R^{-\frac{1}{2}} + (\bar{Y}_t^{ob} - \bar{Y}_c^{ob} + \bar{Y}_c^{un}) \cdot \pi_R^{\frac{1}{2}} - \bar{Y}_c^{un} \cdot \alpha\pi_R^{\frac{1}{2}} - (\bar{Y}_c^{un} + \delta^{\#}) \cdot \pi_R^{-\frac{1}{2}} \right]$$

(11)

(Proof in Appendix).

Considering the fact that $\sigma_t^2, \sigma_c^2, \delta^{\#}$ are predetermined and $n_t^{ob}, n_c^{ob}, \bar{Y}_t^{ob}, \bar{Y}_c^{ob}$ are defined by the observed sample, the probit link of the PEV can be explicitly expressed as a function of α and $\pi_R$, conditional on an assumed value of $\bar{Y}_c^{un}$. We will draw on this feature to seek answers for fundamental questions about the robustness of an inference, such as "what does the treatment effect estimated in the unobserved sample have to be such that the PEV is smaller than a desired value for Borman et al. (2008), holding the belief that the proportion of volunteer schools (or schools similar to them) is 0.5 in the target population?" [1]or "what does the proportion of volunteer schools (and schools similar to them) have to be such that the PEV is smaller than a desired value for Borman et al. (2008), when the treatment effect in the unobserved sample is zero?". Once the values of parameters $n_t^{ob}, n_c^{ob}, \sigma_t^2, \sigma_c^2, \bar{Y}_c^{un}, \bar{Y}_t^{ob}, \bar{Y}_c^{ob}$ and the threshold $\delta^{\#}$ are chosen by the researcher, the first question can be answered by pinpointing a threshold of α that makes the PEV just below the desired value. Likewise, the second question can be answered by identifying a threshold of $\pi_R$ that makes the PEV just below the desired value.

## The Bayesian approach

---

Assuming $\bar{Y}_c^{un}$ equals the grand mean of the observed sample.

The Bayesian approach does not differ much from the frequentist approach except that it frames the unobserved sample as a prior distribution and combines it with a likelihood from the observed sample to generate a posterior distribution (Li, 2018). In other words, the frequentist approach is a direct way of computing the sufficient ideal sample statistics from the known observed sample statistics and the conceptualized unobserved sample statistics, whereas the Bayesian approach is an indirect way which constructs a prior based on the unobserved sample and the likelihood based on the observed sample, as has been discussed by Frank & Min (2007).

**Theorem 3:** Suppose we have the following priors and likelihoods for the treated outcome and control outcome:

$$\begin{aligned} &\mu_t \sim N(\bar{Y}_t^{un}, \frac{\sigma_t^2}{n_t^{un}}) \quad \mu_c \sim N(\bar{Y}_c^{un}, \frac{\sigma_c^2}{n_c^{un}}) \\ &Y_t \sim N(\mu_t, \sigma_t^2) \qquad Y_c \sim N(\mu_c, \sigma_c^2) \end{aligned} \tag{12}$$

and we further assume the treated outcome and control outcome are independent. The posterior distribution of $\delta$ conditional on the observed sample is identical to the distribution in (7). That is, we have:

$$\delta \,|\, \mathbf{D^{ob}} \sim N(\theta_t - \theta_c, \phi_t + \phi_c) \tag{13}$$

where $\theta_t, \theta_c, \phi_t, \phi_c$ are as specified in (8).

(Proof in Appendix).

Theorem 3 reveals that the frequentist approach and the Bayesian approach lead to the same inferential distribution for $\delta$, provided that priors and likelihoods are in the form of (5.6).

The functional form of (12) implies two assumptions: First, I assume normality for the distributions of treated outcome and control outcome. Second, I assume the variances of these two distributions are given since only the means are parameters for the simple estimator. Theorem 3 allows us to interpret the posterior distribution of $\delta$ as its distribution based on the ideal sample, which was a key observation of Frank & Min (2007). Hoff (2009) (also see Diaconis & Ylvisaker, 1979, 1985) has pointed out that such an interpretation is generically possible for a likelihood belongs to the exponential family and its conjugate prior.

As in the frequentist approach, we can simplify the expressions of $\theta_t, \theta_c, \phi_t, \phi_c$ by introducing $\pi_R$ in the Bayesian approach, obtaining the same expressions as in (9). Needless to say, the PEV in the Bayesian approach is the same function of $\alpha$ and $\pi_R$ as described in theorem 2, and one would follow the same procedure to locate the thresholds for either of them.

**$\delta^{\#}$ as a statistical threshold**

Deciding $\delta^{\#}$ is a necessary step for identifying the probit function in (10) or (11). As a threshold of claiming an effect is found, $\delta^{\#}$ could be purely non-statistical and set based on transaction cost and/or policy implications (See Frank et al. (2013) for a discussion). From a standpoint of hypothesis testing, $\delta^{\#}$ is most likely a statistical choice and is commonly computed as $\pm 1.96 * se_{\hat{\delta}^{id}}$ associated with the level of significance 0.05. One pitfall in deriving such statistical threshold is finding $se_{\hat{\delta}^{id}}$, which is the standard error of the simple estimator of average treatment effect based on the ideal sample.

Note that $se_{\hat{\delta}^{id}}$ for the simple estimator is:

$$se_{\hat{\delta}^{id}} = \sqrt{\phi_t + \phi_c} = \sqrt{\pi_R \left( \frac{\sigma_t^2}{n_t^{ob}} + \frac{\sigma_c^2}{n_c^{ob}} \right)} \qquad (14)$$

Correspondingly, the probit functions in (10) and (11) are transformed as follows:

When a significant positive effect has been concluded, (10) is updated as:

$$probit(PEV) = \frac{1}{\sqrt{\frac{\sigma_t^2}{n_t^{ob}} + \frac{\sigma_c^2}{n_c^{ob}}}} \left[ \bar{Y}_c^{un} \cdot \alpha\pi_R^{\frac{1}{2}} - \bar{Y}_c^{un} \cdot \alpha\pi_R^{-\frac{1}{2}} - (\bar{Y}_t^{ob} - \bar{Y}_c^{ob} + \bar{Y}_c^{un}) \cdot \pi_R^{\frac{1}{2}} + \bar{Y}_c^{un} \cdot \pi_R^{-\frac{1}{2}} \right] + 1.96$$

(15)

When a significant negative effect has been concluded, (11) is updated as:

$$probit(PEV) = \frac{1}{\sqrt{\frac{\sigma_t^2}{n_t^{ob}} + \frac{\sigma_c^2}{n_c^{ob}}}} \left[ \bar{Y}_c^{un} \cdot \alpha\pi_R^{-\frac{1}{2}} + (\bar{Y}_t^{ob} - \bar{Y}_c^{ob} + \bar{Y}_c^{un}) \cdot \pi_R^{\frac{1}{2}} - \bar{Y}_c^{un} \cdot \alpha\pi_R^{\frac{1}{2}} - \bar{Y}_c^{un} \pi_R^{-\frac{1}{2}} \right] + 1.96$$

(16)

## The relationship between the PEV and unobserved sample for the regression estimator

### The frequentist approach

The main difference between the regression estimator and the simple estimator is the structure of the ideal sample $\mathbf{D^{id}}$. We have learned that the ideal sample consists of the outcomes Y only whereas the ideal sample for the regression estimator consists of observations which contain not only the outcome Y but also the treatment indicator W and the covariates $Z_1$, …, $Z_p$.. Keeping such data structure in mind, we have the following theorem:

**Theorem 4:** Suppose the unobserved sample is obtained and the residual variance $\sigma^2$ for regression is given. Under the classical linear regression model (CLRM), the distribution of $\delta$ , which in this case represents the true regression coefficient of the treatment indicator W (i.e., $\beta_W$ ), will be as follows conditional on the ideal sample:

$$\delta \mid \mathbf{D^{id}} \sim N\left(\frac{\hat{\sigma}_{WY}^{id} - \mathbf{S_{WZ}^{id}}(\mathbf{S_{ZZ}^{id}})^{\mathbf{-1}}\mathbf{S_{ZY}^{id}}}{\hat{\sigma}_{WW}^{id} - \mathbf{S_{WZ}^{id}}(\mathbf{S_{ZZ}^{id}})^{\mathbf{-1}}\mathbf{S_{ZW}^{id}}}, \frac{\sigma^2}{n^{un}+n^{ob}}(\hat{\sigma}_{WW}^{id} - \mathbf{S_{WZ}^{id}}(\mathbf{S_{ZZ}^{id}})^{\mathbf{-1}}\mathbf{S_{ZW}^{id}})^{\mathbf{-1}}\right)$$

(17)

where:

$$\hat{\sigma}_{Z_iZ_j}^{id} = \lambda\hat{\sigma}_{Z_iZ_j}^{un} + (1-\lambda)\hat{\sigma}_{Z_iZ_j}^{ob} + (1-\lambda)\lambda(\bar{Z}_i^{ob} - \bar{Z}_i^{un})(\bar{Z}_j^{ob} - \bar{Z}_j^{un}) \text{ for } i \neq j = 1,2,\ldots,p$$
$$\hat{\sigma}_{WZ_i}^{id} = \lambda\hat{\sigma}_{WZ_i}^{un} + (1-\lambda)\hat{\sigma}_{WZ_i}^{ob} + (1-\lambda)\lambda(\bar{W}^{ob} - \bar{W}^{un})(\bar{Z}_i^{ob} - \bar{Z}_i^{un}) \text{ for } i = 1,2,\ldots,p$$
$$\hat{\sigma}_{WW}^{id} = \lambda\hat{\sigma}_{WW}^{un} + (1-\lambda)\hat{\sigma}_{WW}^{ob} + (1-\lambda)\lambda(\bar{W}^{ob} - \bar{W}^{un})^2$$
$$\hat{\sigma}_{Z_iY}^{id} = \lambda\hat{\sigma}_{Z_iY}^{un} + (1-\lambda)\hat{\sigma}_{Z_iY}^{ob} + (1-\lambda)\lambda(\bar{Z}_i^{ob} - \bar{Z}_i^{un})(\bar{Y}^{ob} - \bar{Y}^{un}) \text{ for } i = 1,2,\ldots,p$$
$$\hat{\sigma}_{WY}^{id} = \lambda\hat{\sigma}_{WY}^{un} + (1-\lambda)\hat{\sigma}_{WY}^{ob} + (1-\lambda)\lambda(\bar{W}^{ob} - \bar{W}^{un})(\bar{Y}^{ob} - \bar{Y}^{un})$$

(18)

and:

$$\lambda = \frac{n^{un}}{n^{un}+n^{ob}} \qquad (19)$$

(Proof in Appendix).

Theorem 4 shows that one can compute the PEV over the distribution shown in (17) for the regression estimator and suggests that the PEV is related to the unobserved sample means, variances, covariances and sample size. Correspondingly, Theorem 5 explicitly describes the relationship between the PEV and the unobserved sample statistics as follows:

**Theorem 5:** Drawing on the distribution (17), the probit link of the PEV is a function of the ideal sample variances, the ideal sample covariances, and the ideal sample size (i.e., $n^{un} + n^{ob}$ ). Specifically, when a significant positive effect has been concluded, we have:

$$probit(PEV) = \frac{\sqrt{n^{un}+n^{ob}}}{\sigma\sqrt{\hat{\sigma}^{id}_{WW} - \mathbf{S}^{\mathbf{id}}_{\mathbf{WZ}}(\mathbf{S}^{\mathbf{id}}_{\mathbf{ZZ}})^{-1}\mathbf{S}^{\mathbf{id}}_{\mathbf{ZW}}}}[\delta^{\#}(\hat{\sigma}^{id}_{WW} - \mathbf{S}^{\mathbf{id}}_{\mathbf{WZ}}(\mathbf{S}^{\mathbf{id}}_{\mathbf{ZZ}})^{-1}\mathbf{S}^{\mathbf{id}}_{\mathbf{ZW}}) - (\hat{\sigma}^{id}_{WY} - \mathbf{S}^{\mathbf{id}}_{\mathbf{WZ}}(\mathbf{S}^{\mathbf{id}}_{\mathbf{ZZ}})^{-1}\mathbf{S}^{\mathbf{id}}_{\mathbf{ZY}})]$$

(20)

Or when a significant negative effect has been concluded, we have:

$$probit(PEV) = \frac{\sqrt{n^{un}+n^{ob}}}{\sigma\sqrt{\hat{\sigma}^{id}_{WW} - \mathbf{S}^{\mathbf{id}}_{\mathbf{WZ}}(\mathbf{S}^{\mathbf{id}}_{\mathbf{ZZ}})^{-1}\mathbf{S}^{\mathbf{id}}_{\mathbf{ZW}}}}[(\hat{\sigma}^{id}_{WY} - \mathbf{S}^{\mathbf{id}}_{\mathbf{WZ}}(\mathbf{S}^{\mathbf{id}}_{\mathbf{ZZ}})^{-1}\mathbf{S}^{\mathbf{id}}_{\mathbf{ZY}}) - \delta^{\#}(\hat{\sigma}^{id}_{WW} - \mathbf{S}^{\mathbf{id}}_{\mathbf{WZ}}(\mathbf{S}^{\mathbf{id}}_{\mathbf{ZZ}})^{-1}\mathbf{S}^{\mathbf{id}}_{\mathbf{ZW}})]$$

(21)

(Proof in Appendix).

The essence of theorem 5 is that the PEV is a function of the unobserved sample statistics, as one can always express the ideal sample statistics in (20) or (21) as functions of the observed sample statistics which are known and the unobserved sample statistics which are unknown parameters, according to (18). Just like the simple estimator case, for the regression estimator, we can utilize theorem 5 to find the threshold of a focal unobserved sample statistic corresponding to a desired value of the PEV, while holding all other unobserved sample statistics fixed. For example, to quantify how many more classrooms from non-volunteer schools are needed (assuming the correlation between OCR curriculum and reading scores in sampled non-volunteer classrooms is zero), the answer is provided by the threshold of $n^{un}$ (the unobserved sample size) conditional on the assumption $\hat{\sigma}_{WY}^{un} = 0$ for the PEV to be lower than a desired value. The analytical strategy will be systematized and illustrated in the example section.

**The Bayesian approach**

In this section, I show that the distribution (17) can be derived through a Bayesian model. The logic is the same as the Bayesian approach for the simple estimator: the prior should be built from the unobserved sample and likelihood should be built from the observed sample (Li, 2018). Following this logic, we have theorem 6:

**Theorem 6:** Conditional on the following prior and likelihood:

$$\begin{aligned} &\boldsymbol{\beta} \sim N(((\mathbf{X^{un}})^{\mathbf{T}}\mathbf{X^{un}})^{\mathbf{-1}}(\mathbf{X^{un}})^{\mathbf{T}}\mathbf{Y^{un}}, \sigma^2((\mathbf{X^{un}})^{\mathbf{T}}\mathbf{X^{un}})^{\mathbf{-1}}) \\ &Y_i \mid \boldsymbol{\beta}, \mathbf{X_i^{ob}} \sim N(\mathbf{X_i^{ob}}\boldsymbol{\beta}, \sigma^2) \end{aligned} \tag{22}$$

the posterior distribution of $\delta$ will be identical to (17).

(Proof in Appendix).

In theorem 6, $\mathbf{X_i}$ is a vector contains 1 and values of W, $Z_1$, …, $Z_p$ for the $i^{th}$ individual. Likewise, $\mathbf{X}$ is the matrix whose $i^{th}$ row is $\mathbf{X_i}$, and both $\mathbf{X}$ and $\mathbf{X_i}$ have a superscript attached (to signal which sample it belongs to). $\boldsymbol{\beta}$ refers to the vector of regression coefficients. Theorem 6 draws on two assumptions: First, like the frequentist approach, we need to assume CLRM for the regression of Y on X, that is, the residuals are normally distributed with a common variance $\sigma^2$. Second, $\sigma^2$ is a given value as only the regression coefficients are considered as parameters in the Bayesian model (22). Conceptually, the prior in (22) can be thought of as the distribution of **β** based on the unobserved sample and the posterior can be thought of as the distribution of **β** based on the ideal sample (see my proof if interested).

### $\delta^{\#}$ as a statistical threshold

As discussed earlier, the threshold $\delta^{\#}$ is used for deciding whether a significant effect is found and could be prespecified by researchers. It is conventionally statistical, from a hypothesis testing perspective. Consistent with the previous discussion, the statistical threshold $\delta^{\#}$ is $\pm 1.96 * se_{\hat{\delta}^{id}}$ for the regression estimator. However, the formula of computing $se_{\hat{\delta}^{id}}$ is different from section 5 because $\hat{\delta}^{id}$ now becomes the estimated regression coefficient of the treatment indicator W based on the ideal sample.

The formula of computing the $se_{\hat{\delta}^{id}}$ for the regression estimator is provided below:

$$se_{\hat{\delta}^{id}} = \sqrt{\frac{\sigma^2}{n^{un} + n^{ob}} (\hat{\sigma}_{WW}^{id} - \mathbf{S}_{\mathbf{WZ}}^{\mathbf{id}} (\mathbf{S}_{\mathbf{ZZ}}^{\mathbf{id}})^{-1} \mathbf{S}_{\mathbf{ZW}}^{\mathbf{id}})^{-1}} \tag{23}$$

Calculating $\delta^{\#}$ as $1.96 * se_{\hat{\delta}^{id}}$ for a reported significant positive effect and plugging in this threshold into (20) will update the probit function as (24):

$$probit(PEV) = 1.96 - \frac{\sqrt{n^{un} + n^{ob}}}{\sigma \sqrt{\hat{\sigma}_{WW}^{id} - \mathbf{S}_{\mathbf{WZ}}^{\mathbf{id}} (\mathbf{S}_{\mathbf{ZZ}}^{\mathbf{id}})^{-1} \mathbf{S}_{\mathbf{ZW}}^{\mathbf{id}}}} (\hat{\sigma}_{WY}^{id} - \mathbf{S}_{\mathbf{WZ}}^{\mathbf{id}} (\mathbf{S}_{\mathbf{ZZ}}^{\mathbf{id}})^{-1} \mathbf{S}_{\mathbf{ZY}}^{\mathbf{id}}) \tag{24}$$

Otherwise, calculating $\delta^{\#}$ as $-1.96 * se_{\hat{\delta}^{id}}$ for a reported significant negative effect and plugging this threshold into (21) will update this probit function as (25):

$$probit(PEV) = 1.96 + \frac{\sqrt{n^{un} + n^{ob}}}{\sigma\sqrt{\hat{\sigma}_{WW}^{id} - \mathbf{S}_{\mathbf{WZ}}^{\mathbf{id}}(\mathbf{S}_{\mathbf{ZZ}}^{\mathbf{id}})^{-1}\mathbf{S}_{\mathbf{ZW}}^{\mathbf{id}}}}(\hat{\sigma}_{WY}^{id} - \mathbf{S}_{\mathbf{WZ}}^{\mathbf{id}}(\mathbf{S}_{\mathbf{ZZ}}^{\mathbf{id}})^{-1}\mathbf{S}_{\mathbf{ZY}}^{\mathbf{id}}) \quad (25)$$

## Example: The effect of Open Court Reading curriculum on reading achievement

### Overview

The Open Court Reading (OCR) program is a curriculum that is rooted in research-based practices and has been in the market for a long time and widely adopted by many districts and schools. Although OCR is potentially a beneficial program because it responds to recommendations from research that focused on developing early reading skills, its effect had initially not been assessed and confirmed by a randomized experiment. Seeing this, Borman et al. (2008) designed a multisite cluster randomized experiment and randomly drew 6 schools from the schools had contacted and shown their interest to SRA/McGraw Hill, the publisher of the OCR curriculum. Those 6 schools came from six different states (Florida, Georgia, Idaho, Indiana, North Carolina and Texas) and were geographically, ethnically and socioeconomically representative of schools in US. Subsequently, Borman et al. defined a block as a single grade of one sampled school and within each block classrooms were randomly assigned to the OCR group or the control group (business as usual). The final sample of Borman et al. (2008) included five schools (the Georgia school dropped out) and 49 classrooms of which 27 classrooms were assigned to the OCR group. Controlling for the pretest scores and block membership, Borman et al. (2008) estimated the effect of OCR as 7.95 (on reading composite scores) which was statistically significant and went on to conclude that “the outcomes from these analyses provided not only evidence of the promising 1-year effects of OCR on students' reading outcomes but also suggest that these effects may be replicated across varying contexts with rather consistent and positive results".

Ideally, the findings of Borman et al. (2008) imply that, the estimated effect of OCR would be around 7.95 if one were to conduct a large-scale completely randomized experiment elsewhere or collected a nationally representative sample. At the very least, one should be able to conclude a significantly positive effect of OCR if the experiment were replicated. However, such claims are not necessarily warranted as the external validity of Borman et al. (2008) is debatable, given all sampled schools of Borman et al. (2008) were drawn from the volunteer schools. Specifically, Frank et al. (2013) note that the effect of the OCR program might not have been as large as the one reported by Borman et al. (2008) had their study been conducted in non-volunteer schools, possibly because volunteer schools might have anticipated the OCR program was highly effective for them given their student composition. If this were the case, the effect of the OCR curriculum would be overestimated and the inference drawn by Borman et al. (2008) may not be valid for all targeted schools.

Following the above argument, the external validity of Borman et al. (2008) is evaluated next by quantifying its robustness as the PEV proposed in this paper. Particularly, the analytical procedure includes nine steps: 1-choose a focal parameter, 2-specify the parametric values, 3-determine the decision threshold, 4-obtain the probit model, 5-specify a desired value of the PEV, 6-derive the inequality for the focal parameter, 7-locate the threshold of the focal parameter, 8-conduct auxiliary analysis, 9-repeat step 5 through step 8 for different desired values of the PEV. The flowchart of this analytical procedure is displayed below:

(Insert Figure 2 here)

**The simple estimator**

1-Choose a focal parameter: Technically speaking, all the symbols appear in (10) (or (15) when $\delta^{\#}$ is the statistical threshold) are the parameters in our context. However, some of them are observed sample statistics and they are naturally fixed. The variance terms are given by assumption. Most importantly, the focal parameters should quantify concerns about the robustness of the inference directly. For example, in Borman et al. (2008), the inquiry about its external validity could be phrased as follows: "to invalidate their inference, what would the mean reading scores of the OCR and the control groups in non-volunteer schools need to be if we sample more classrooms from the non-volunteer schools" or "to invalidate their inference, how many more classrooms from non-volunteer schools can be drawn and added to the observed sample when there is no treatment effect in those unobserved classrooms". Obviously, the first question can be addressed by α and the second question can be addressed by $\pi_R$. For other studies with debatable external validity, α and $\pi_R$ will also likely be focal parameters for the simple estimator.

2-Specify the parametric values: Considering the focal parameters are set as α and $\pi_R$, we will need to specify values of other terms in the probit model (10) or (15). According to Borman et al. (2008) and Frank et al. (2013), they are set as follows:

$$
\begin{aligned}
\bar{Y}_c^{un} &= 611.5 \\
\bar{Y}_t^{ob} &= 615 \\
\bar{Y}_c^{ob} &= 607 \\
\sigma_t^2 &= 45 \\
\sigma_c^2 &= 45 \\
n_t^{ob} &= 27 \\
n_c^{ob} &= 22
\end{aligned}
\tag{26}
$$

In (26), the unobserved control sample mean is set as 611.5, which is the mean of whole observed sample and mirrors the null hypothesis: $\delta = 0$.

3-Determine the decision threshold: Before finalizing the probit model, one also needs to determine the value of $\delta^{\#}$ which is the threshold of concluding a significant positive effect. As discussed earlier, $\delta^{\#}$ could be a fixed number or a threshold for statistical significance, i.e., $1.96 * se_{\hat{\delta}^{id}}$. In this case, I choose $\delta^{\#}$ to be the statistical threshold.

4-Obtain the probit model: Guided by (15), the resultant probit model is as follows:

$$
probit(PEV) = 317.38\alpha\pi_R^{\frac{1}{2}} - 317.38\alpha\pi_R^{-\frac{1}{2}} - 321.54\pi_R^{\frac{1}{2}} + 317.38 \cdot \pi_R^{-\frac{1}{2}} + 1.96 \tag{27}
$$

5-Specify a desired value of the PEV: The purpose of doing this is to derive an inequality and identify a threshold for a focal parameter. I will show later (in the ninth step) that users should choose different desired values of the PEV in order to gain comprehensive knowledge about the robustness of external validity. For illustrative purpose I set the desired value of the PEV temporarily as 0.5.

6-Derive the inequality for the focal parameter: To make the PEV smaller than 0.5, I use the following quadratic inequality to identify the threshold of $\pi_R$:

$$(317.38\alpha - 321.54)\pi_R + 1.96\pi_R^{0.5} + (317.38 - 317.38\alpha) < 0 \tag{28}$$

Likewise, I use the following linear inequality to identify the threshold of α for the PEV to be smaller than 0.5:

$$\alpha > \frac{321.54\pi_R^{0.5} - 317.38\pi_R^{-0.5} - 1.96}{317.38(\pi_R^{0.5} - \pi_R^{-0.5})} \tag{29}$$

It is clear that the inequality for identifying the threshold of $\pi_R$ needs to be conditional on a value of α, and vice versa.

7-Locate the threshold of the focal parameter: From (28), $\pi_R$ should be larger than 0.2228 such that the PEV is smaller than 0.5 conditional on α=1. From (29), α should be larger than 0.9966 such that the PEV is smaller than 0.5 conditional on $\pi_R = 0.46$. Here, the condition α=1 represents the null hypothesis: $\delta = 0$ and the condition $\pi_R = 0.46$ is the previously proposed threshold of $\pi_R$ to invalidate Borman et al. (2008)'s inference. (Frank et al. (2013).

8-Conduct auxiliary analysis: Although in the seventh step we derived the thresholds of α and $\pi_R$ with regard to the PEV = 0.5, those thresholds have little meaning in an empirical context. Given α has to be larger than 0.9966, we can further calculate the threshold of the unobserved treated sample mean as well as the threshold of average treatment effect in an unobserved sample. The threshold of the unobserved treated sample mean (i.e., $\bar{Y}_t^{un}$) should be 611.5*0.9966 = 609.42 and the threshold of average treatment effect in the unobserved sample should be 611.5*(0.9966-1) = -2.08. These thresholds inform that the mean reading score of the OCR classrooms sampled from the non-volunteer schools should be larger than 609.42 if the mean reading score of the controlled classrooms were 611.5 and resultantly the estimated average treatment effect in the unobserved sample should be larger than -2.08, in order to make the PEV lower than 0.5. Drawing on these thresholds, the threshold of estimated treatment effect based on the ideal sample (i.e., $\hat{\delta}^{id}$ , which equals $\theta_t - \theta_c$ ) is computed as 8*0.46 -2.08*0.54 = 2.56. This means the estimated treatment effect of the OCR program based on the ideal sample should be larger than 2.56 in order to make the PEV smaller than 0.5, should the ideal sample become available.

The threshold of $\pi_R$ can be interpreted as follows: Assuming there is no treatment effect for the OCR curriculum in non-volunteer schools, we will need the observed sample of Borman et al. (2008) to represent at least 22.3% of the entire U.S. schools conditional on the parametric values in (26) so that the PEV is smaller than 0.5. Drawing on this threshold of $\pi_R$ , we can directly compute the threshold of $\hat{\delta}^{id}$ as 8*0.2228 = 1.78. This shows the estimated treatment effect of OCR should be larger than 1.78 based on an ideal sample for the PEV to be lower than 0.5.

9-Repeat the step 5 through the step 8 for different desired values of the PEV: The thresholds of $\pi_R$ and α can be repeatedly calculated for different desired values of the PEV based on the same conditions. Table 1 provides the thresholds of α (as well as thresholds of the unobserved treated sample mean and estimated average treatment effect in the ideal sample) when $\pi_R = 0.46$ for the PEV = 0.1, 0.2, …, 0.9. Table 2 provides the threshold of $\pi_R$ (as well as thresholds of $\hat{\delta}^{id}$ ) when α = 1 for the PEV = 0.1, 0.2, …, 0.9. The general pattern is: As the external validity of Borman et al. (2008) becomes stronger (signified by a lower PEV), the thresholds of both α and $\pi_R$ should increase, which corresponds to a larger estimated treatment effect based on an ideal sample. In other words, the PEV is negatively related to the estimated treatment effect based on the ideal sample.

(Insert Table 1 here)

(Insert Table 2 here)

**The regression estimator**

For the regression estimator, the external validity of Borman et al. (2008) is assessed through the same procedure as for the simple estimator, except that the pretest is treated as a covariate and the average treatment effect is estimated as the regression coefficient of the treatment indicator of OCR:

1-Choose a focal parameter: As for the simple estimator, theoretically any symbol(s) in (20) can be focal parameter(s) but the observed sample statistics, residual variance and decision threshold $\delta^{\#}$ are typically fixed. In this example, our inquiry about the external validity of Borman et al. (2008) is "to invalidate the inference, how many more classrooms in the non-volunteer schools are needed when the correlation between OCR and posttest reading scores is zero in non-volunteer schools". The above condition reflects the null hypothesis: $\delta = 0$. To answer this question, the focal parameter is chosen as $n^{un}$ , which denotes the unobserved sample size.

2-Specify the parametric values: They are specified as follows (Z: pretest, W: OCR treatment status, Y: posttest):

$$
\begin{aligned}
&\hat{\sigma}_{WY}^{un} = 0,\ \hat{\sigma}_{WY}^{ob} = 2.33 \\
&n^{ob} = 49, \sigma^2 = 32 \\
&\hat{\sigma}_{ZZ}^{un} = \hat{\sigma}_{ZZ}^{ob} = 2079.36 \\
&\hat{\sigma}_{ZY}^{un} = \hat{\sigma}_{ZY}^{ob} = 1832.2 \\
&\hat{\sigma}_{WW}^{un} = \hat{\sigma}_{WW}^{ob} = 0.25 \\
&\hat{\sigma}_{ZW}^{un} = \hat{\sigma}_{ZW}^{ob} = 0.39 \\
&\bar{Z}^{un} = \bar{Z}^{ob} = 576.62 \\
&\bar{Y}^{un} = \bar{Y}^{ob} = 609.96 \\
&\bar{W}^{un} = \bar{W}^{ob} = 0.55
\end{aligned}
\tag{30}
$$

3-Determine the decision threshold: In this example, I choose to use the statistical threshold offered by (23).

4-Obtain the probit model: Directed by (24) and the above parametric values, the probit model should be written as follows:

$$
probit(PEV) = 1.96 - \frac{40.36}{\sqrt{n^{un} + 49}} + 0.12\sqrt{n^{un} + 49} \tag{31}
$$

5-Specify a desired value of the PEV: Again, it is not necessary to fix the PEV, but for illustrative purposes I set the desired value of the PEV to 0.5.

6-Derive the inequality for the focal parameter: For the PEV to be smaller than 0.5, the inequality is straightforward from (31):

$$
1.96 - \frac{40.36}{\sqrt{n^{un} + 49}} + 0.12\sqrt{n^{un} + 49} < 0 \tag{32}
$$

7-Locate the threshold of the focal parameter: The threshold of $n^{un}$ is 91, which means one can at most add 91 classrooms sampled from non-volunteer schools to the observed sample while keeping the PEV lower than 0.5, assuming the correlation between OCR and posttest reading scores in non-volunteer schools is zero. Of course, this threshold value also depends on the parametric values specified in (30).

8-Conduct auxiliary analysis: We can further derive the estimated treatment effect based on the ideal sample as follows:

$$\hat{\delta}^{id} = \frac{\hat{\sigma}_{WY}^{id} - \mathbf{S}_{\mathbf{WZ}}^{\mathbf{id}}(\mathbf{S}_{\mathbf{ZZ}}^{\mathbf{id}})^{-1}\mathbf{S}_{\mathbf{ZY}}^{\mathbf{id}}}{\hat{\sigma}_{WW}^{id} - \mathbf{S}_{\mathbf{WZ}}^{\mathbf{id}}(\mathbf{S}_{\mathbf{ZZ}}^{\mathbf{id}})^{-1}\mathbf{S}_{\mathbf{ZW}}^{\mathbf{id}}} \quad (33)$$

The computations of the ideal sample variances and covariances are guided by (18). The threshold of $n^{un} = 91$ corresponds to the estimated regression coefficient of W based on the ideal sample is 1.89, which means the estimated average treatment effect of OCR should be larger than 1.89 in order to make the PEV lower than 0.5 in this case.

9- Repeat the step 5 through the step 8 for different desired values of the PEV: The threshold of $n^{un}$ as well as the corresponding threshold of $\hat{\delta}^{id}$ are repeatedly calculated for the PEV = 0.1, 0.2, …, 0.9. Those thresholds are tabulated in table 3. Like the simple estimator, for the regression estimator we observe a similar general pattern: as the external validity of Borman et al. (2008) grows stronger (signified by a lower PEV), the threshold of $n^{un}$ decreases (which indicates the threshold of $\pi_R$ should increase), which corresponds to a larger estimated treatment effect of OCR based on the ideal sample. The PEV is also negatively related to $\hat{\delta}^{id}$ in this case.

(Insert Table 3 here)

**The PEV and retesting the null hypothesis**

In this section, I focus on the interpretation of the PEV, particularly in the context of retesting the null hypothesis when the ideal sample is available. The background is this: Assuming one has tested the null hypothesis: $\delta = 0$ versus the alternative hypothesis: $\delta \neq 0$ and concluded that the null hypothesis should be rejected as a significant positive (or negative) effect is found. One then seeks to retest this null hypothesis and reject it with the same finding when he has the ideal sample. An inference is invalidated if one fails to retain the conclusion he has previously made when the same hypothesis is retested for the ideal sample. This means one fails to find a significant positive/negative effect for the ideal sample but found it for the observed sample.

It is noteworthy that the relationship between the PEV and hypothesis retesting for the ideal sample can be explicitly written as follows:

When a significant positive effect has been concluded and $\delta^{\#} = 1.96 * se_{\hat{\delta}^{id}}$ , we have:

$$probit(PEV) = 1.96 - T \tag{34}$$

When a significant negative effect has been concluded and $\delta^{\#} = -1.96 * se_{\hat{\delta}^{id}}$ , we have:

$$probit(PEV) = 1.96 + T \tag{35}$$

Here T denotes the T-ratio pertaining to retesting the null hypothesis, i.e., $T = \frac{\hat{\delta}^{id}}{se_{\hat{\delta}^{id}}}$ . Obviously, (34) informs us why $\hat{\delta}^{id}$ is negatively related to the PEV, as we observed in table 1 through table 3. This is because the T-ratio is negatively related to the PEV but positively related to $\hat{\delta}^{id}$ . Most importantly, we learn from (34) and (35) that the PEV is primarily a function of the T-ratio.

Graphically, the relationship between the PEV and retesting the null hypothesis for the ideal sample is illustrated below in figure 3, for the earlier example where I use the regression estimator of the effect of OCR with $n^{un}$ as focal parameter. The black curves in figure 3 are distributions corresponding to null hypothesis $\delta = 0$ and the red curves are distributions of $\delta$ conditional on a given ideal sample. Figure 3 depicts the same pattern as manifested in Table 3: as the unobserved sample size increases the distribution corresponding to the null hypothesis and the distribution of $\delta$ conditional on an ideal sample converge, and therefore the PEV will increases. The statistical threshold will, in this case, continue dropping as the ideal sample size grows. Thus figure 3 shows that the PEV can be conceptualized as type II error with regard to retesting the null hypothesis: $\delta = 0$ versus the alternative hypothesis: $\delta = \hat{\delta}^{id}$ when the unobserved sample can be randomly drawn from the non-volunteer schools and merged to the observed sample of Borman et al. (2008). Therefore, assessing external validity through the PEV is linked to power analysis, or more formally, one minus the PEV is the statistical power of testing the null hypothesis $\delta = 0$ versus the alternative hypothesis: $\delta = \hat{\delta}^{id}$ .

(Insert Figure 3 here)

**Make a judgment about external validity**

I have shown how to implement the analytical procedure to identify the thresholds of focal parameter(s) for desired values of the PEV. Given outputs such as in table 1 or table 3, a judgement about external validity should be possible, depending on one's knowledge on what the unobserved sample could be. In general, the rule of thumb in setting up a cutoff value for the PEV is the same as the rule of thumb for statistical power. For example, the cutoff value of the PEV could be set as 0.2 which corresponds to a statistical power equals 0.8 (Cohen, 1988, 1992).

Taking table 1 as an example, if we believe OCR can do no harm (therefore α should be no smaller than 1), then the PEV should be smaller than 0.2 conditional on $\pi_R = 0.46$. Furthermore, this means the estimated average treatment effect of OCR should be larger than 3.65 for the ideal sample. Those thresholds help us quantify our belief about the external validity of Borman et al. (2008). A cutoff value can also be set to decide whether external validity is acceptable. For example, if the cutoff value is 0.2 then the external validity of Borman et al. (2008) should be deemed acceptable because we believe the upper bound of the PEV in this case is 0.2. One can also bound the PEV through belief about the unobserved sample. For example, in table 2 if one believes the observed sample of Borman et al. (2008) can at least represent 20% but at most represent 50% of the targeted schools, according to table 2, the PEV should lie in the interval [0.16, 0.54] and the estimated average treatment effect of OCR should lie in the interval [1.6, 4].

## Discussion

### Summary

This article recasts the problem of evaluating external validity as a missing data problem where the unobserved sample is needed in order to form the ideal sample which would legitimatize a claim of strong external validity. The unobserved sample is a complement to the observed sample, but it can only exist in a thought experiment. To formalize such a thought experiment, the definitions of the unobserved sample as well as the non-represented part of target population are presented for the simple estimator and the regression estimator. Drawing on those definitions, the probability of invalidating a causal inference due to limited external validity, i.e., the PEV, is then defined in the context of hypothesis testing provided the null hypothesis has already been rejected for the observed sample. The PEV specifically is concerned with the likelihood of rejecting the null hypothesis again for an ideal sample, should it become accessible. Aiming at the relationship between the PEV and the unobserved sample in a thought experiment, this article offers frequentist as well as Bayesian approaches to determine this relationship. As illustrated by the example of Borman et al. (2008), external validity can be evaluated by characterizing the unobserved samples that would make the PEV lower than a desired value, which will inform the debate about external validity.

The scholarly significance of this study manifests in three aspects: First of all, this study transforms vague beliefs about the unobserved sample into concrete probabilistic index of external validity, i.e., the PEV. As shown earlier, external validity is evaluated through a range of the PEV based on possible values of an unobserved sample statistic of main interest. This helps promote critical thinking as well as scientific debate about external validity as one can express opinions about external validity in the precise terms of a probabilistic index (Li & Frank, 2020a, 2020b). Second, this study demonstrates that under certain conditions the frequentist approach and the Bayesian approach will generate the same result, as evidenced by theorem 3 and theorem 6. Such a feature would appeal to both frequentist thinkers and Bayesian thinkers. Third, the PEV has an intuitive interpretation that is rooted in hypothesis testing. Particularly, the PEV could be seen as type II error associated with testing the null hypothesis: $\delta = 0$ versus the alternative hypothesis: $\delta = \hat{\delta}^{id}$ and thus it reflects the statistical power for the ideal sample.

## Comparisons with other similar indices

Robustness indices in Frank & Min (2007) and Frank et al. (2013): The robustness indices in Frank & Min (2007) focused on what the unobserved sample needs to be so that an inference would be invalidated due to limited external validity. As in this article, their robustness indices emphasize the necessity of conceptualizing the unobserved sample and thus preparing the ideal sample for inference. Frank et al. (2013) has essentially the same framework as Frank & Min (2007) except that it discussed both limited external validity and limited internal validity cases. This study extends the idea of Frank & Min (2007) and Frank et al (2013) in three ways: First, it creates a probabilistic robustness index, i.e., the PEV. Second, it delves deeper into the conceptualization of the unobserved sample. One would be able to ask and answer questions about the statistics that characterize the unobserved sample in quantifying the robustness of external validity. Third, this study generalizes the Bayesian discussion in Frank & Min (2007) and shows the Bayesian paradigm of robustness indices can yield the same results as the frequentist paradigm under certain conditions.

Probability of replication in Killeen (2005): This article is similar to Killeen (2005) as both the PEV and the probability of replication (PR) are rooted in hypothesis testing and intended for the questions of the same nature: what is the probability of replicating the previous significant finding? The main difference between Killeen (2005) and this article is the former has a meta-analytic perspective and the latter has a missing data perspective. Specifically, PR does not require one to conceptualize and characterize the unobserved sample that renders inference based on the ideal sample vulnerable, which is unavoidable in assessing external validity.

Generalizability index in Tipton (2014): The framework of Tipton (2014) is the same as Stuart et al. (2011) and O'Muircheartaigh & Hedges (2014). Their central idea is defining sampling propensity scores based on a key assumption of unconfounded sample selection and comparing sampling propensity scores of observed sample and the target population. The generalizability index in Tipton (2014) quantifies the distance between the two distributions of sampling propensity scores in the population and the sample. The PEV is different from this generalizability index as it does not require the unconfounded sample selection assumption given it directly models unobserved sample and its influence on hypothesis testing. By this logic, the PEV frees researchers from the bias incurred by the unconfounded sample selection assumption and the estimation of sampling propensity scores.

Reproducibility probability in Shao & Chow (2002): The definition of reproducibility probability (RP) in Shao & Chow is similar to the definition of the PEV. According to Shao & Chow (2002), RP is "an estimated power of the future trial using the data from the previous trial(s)". Conceptually, when the future trial is retesting hypothesis in an ideal sample, the PEV should be equivalent to 1-RP. However, considering the PEV as a true equivalent of RP has two issues: First, RP is a posterior predictive probability and the PEV is a posterior probability. Second, RP highlights prediction of future power and does not consider the impact of the unobserved sample on current hypothesis testing.

**Limitations**

Even though the PEV is a useful tool of quantifying the robustness of external validity, it has two major limitations: First, it only addresses bias due to non-random sampling error and cannot inform bias due to other sources like measurement error or violation of SUTVA (Rubin, 1980, 1990). Therefore, the PEV is only appropriate for evaluating external validity when the observed sample is suspected to be non-representative of the whole target population. Second, it focuses on the simple and regression estimator of the average treatment effect, which are arguably restricted in practice. However, it is possible to generalize the framework of the PEV to other more complicated models, such as multilevel models and structural equation models.

## Conclusion

Quantifying the strength of external validity is hardly done by a single index. However, the PEV can nurture the scientific discourse about external validity by critically partitioning the target population into observed and unobserved parts and identifying the unobserved sample that will weaken external validity. Relying on the knowledge on the unobserved part of target population, one can circumscribe the possible forms of the unobserved sample and thus learn how it affects external validity.

*Table 1*: Threshold of α when $\pi_R$ is fixed as 0.46

| Values of the PEV | Threshold of α | Threshold of $\bar{Y}_t^{un}$ | Threshold of $\hat{\delta}^{id}$ |
|---|---|---|---|
| 0.1 | 1.0017 | 612.54 | 4.24 |
| 0.2 | 0.9999 | 611.44 | 3.65 |
| 0.3 | 0.9987 | 610.71 | 3.25 |
| 0.4 | 0.9976 | 610.03 | 2.89 |
| 0.5 | 0.9966 | 609.42 | 2.56 |
| 0.6 | 0.9956 | 608.81 | 2.23 |
| 0.7 | 0.9945 | 608.14 | 1.86 |
| 0.8 | 0.9933 | 607.4 | 1.47 |
| 0.9 | 0.9915 | 606.3 | 0.87 |

*Table 2*: Threshold of $\pi_R$ when α is fixed as 1

| Values of the PEV | Threshold of $\pi_R$ | Threshold of $\hat{\delta}^{id}$ |
|---|---|---|
| 0.1 | 0.6095 | 4.88 |
| 0.2 | 0.4553 | 3.64 |
| 0.3 | 0.358 | 2.86 |
| 0.4 | 0.284 | 2.27 |
| 0.5 | 0.2228 | 1.78 |
| 0.6 | 0.1689 | 1.35 |
| 0.7 | 0.1195 | 0.96 |
| 0.8 | 0.0725 | 0.58 |
| 0.9 | 0.0267 | 0.21 |

*Table 3*: Thresholds of $n^{un}$ assuming $\hat{\sigma}_{WY}^{un} = 0$

| Values of the PEV | Threshold of $n^{un}$ | Threshold of $\hat{\delta}^{id}$ |
|---|---|---|
| 0.1 | 36 | 4.00 |
| 0.2 | 51 | 3.19 |
| 0.3 | 64 | 2.67 |
| 0.4 | 77 | 2.25 |
| 0.5 | 91 | 1.89 |
| 0.6 | 107 | 1.55 |
| 0.7 | 126 | 1.23 |
| 0.8 | 152 | 0.90 |
| 0.9 | 195 | 0.50 |

*Figure 1*. The structure of the target population of Borman et al. (2008)

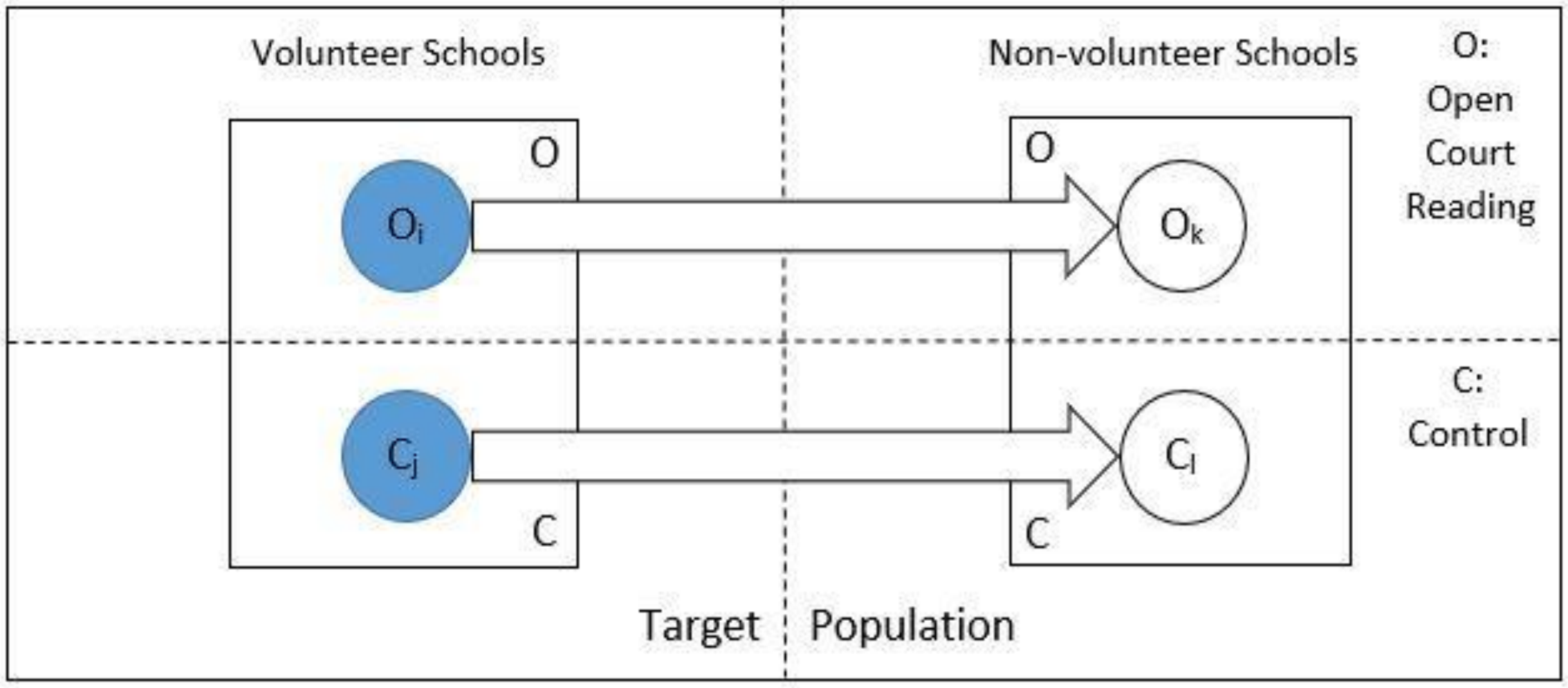

*Figure 2*: Flowchart of the analytical procedure for quantifying the robustness of external validity through the PEV

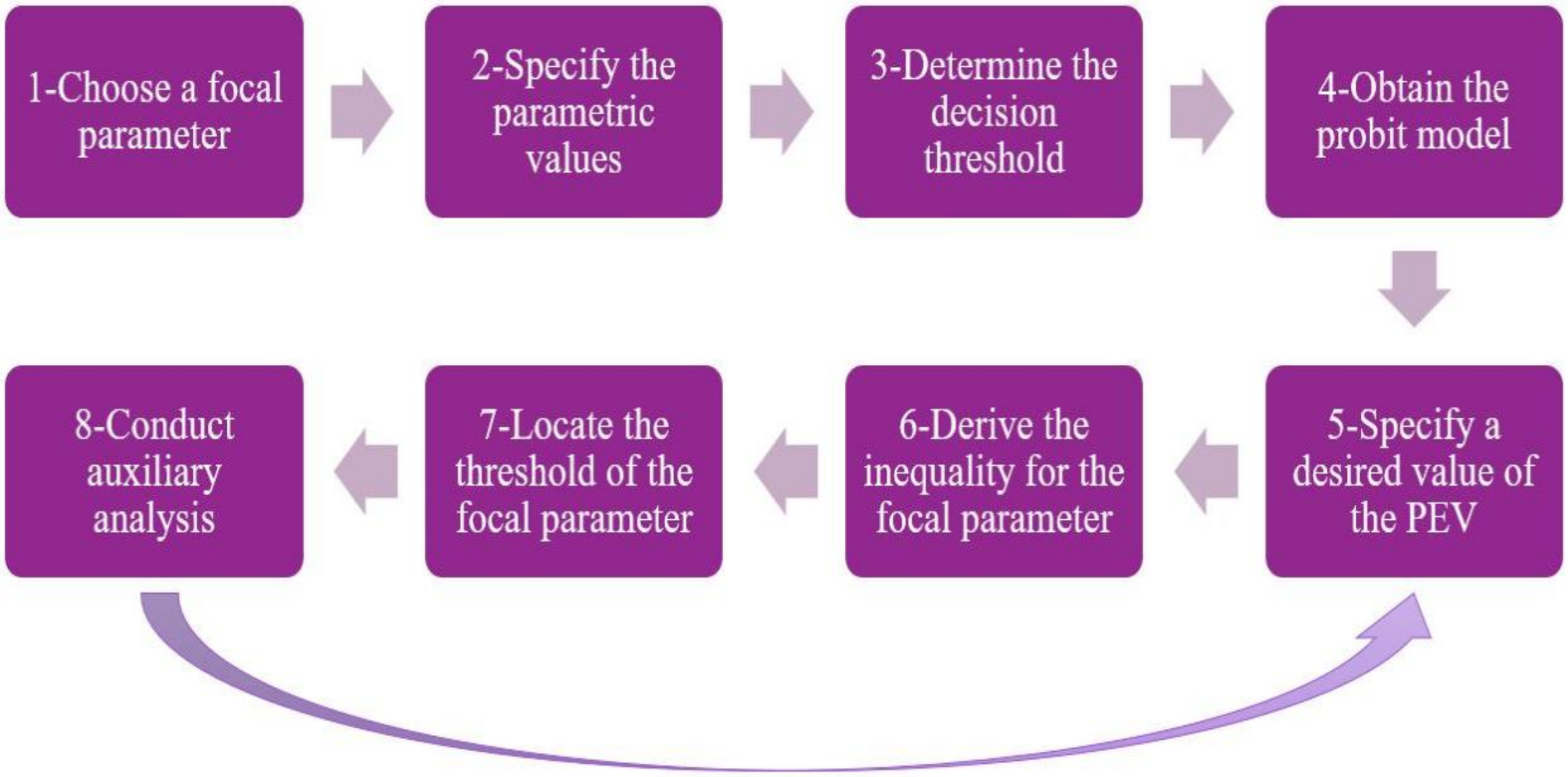

*Figure 3*: The relationship between the PEV and retesting the null hypothesis for the ideal sample for Borman et al. (2008)

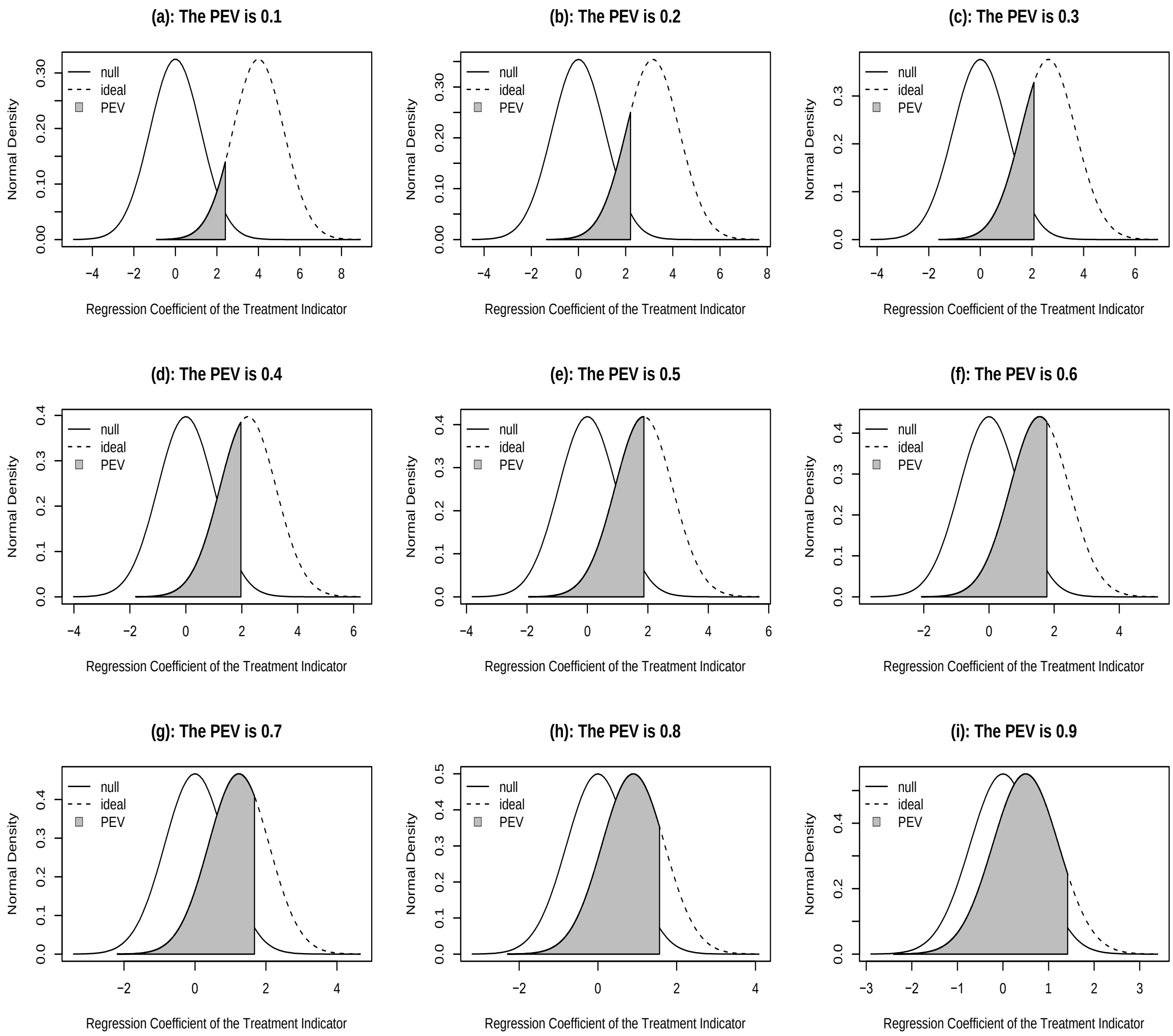

**Appendix**

**Proofs of Theorem 1 Through Theorem 6**

Proof of theorem 1:

First, the distribution of $\delta$ could be derived based on the following pivotal quantity:

$$\frac{\bar{Y}_t^{id} - \bar{Y}_c^{id} - \delta}{\sigma_{\bar{Y}_t^{id} - \bar{Y}_c^{id}}} \sim N(0,1) \tag{A1}$$

The pivotal quantity (A1) is built on the central limit theorem with the belief that the simple estimator $\bar{Y}_t^{id} - \bar{Y}_c^{id}$ should be unbiased for the true average treatment effect $\delta$.

Given the unobserved treated sample size is $n_t^{un}$ and the unobserved treated sample mean is $\bar{Y}_t^{un}$, it is straightforward to write the ideal treated sample mean $\bar{Y}_t^{id}$ as below:

$$\bar{Y}_t^{id} = \frac{n_t^{un}}{n_t^{ob} + n_t^{un}} \bar{Y}_t^{un} + \frac{n_t^{ob}}{n_t^{ob} + n_t^{un}} \bar{Y}_t^{ob} \tag{A2}$$

Similarly, the ideal control sample mean $\bar{Y}_c^{id}$ is written as follows:

$$\bar{Y}_c^{id} = \frac{n_c^{un}}{n_c^{ob} + n_c^{un}} \bar{Y}_c^{un} + \frac{n_c^{ob}}{n_c^{ob} + n_c^{un}} \bar{Y}_c^{ob} \tag{A3}$$

The standard deviation associated with the simple estimator is derived as below:

$$\sigma_{\bar{Y}_t^{id} - \bar{Y}_c^{id}} = \frac{\sigma_t^2}{n_t^{ob} + n_t^{un}} + \frac{\sigma_c^2}{n_c^{ob} + n_c^{un}} \tag{A4}$$

since the variances of the treated and the control outcomes are given. Finally, we can just plug (A2), (A3) and (A4) in (A1), which yields the distribution (3) whose mean and variance are defined by (4).

Proof of theorem 2:

When a significant positive effect has been concluded, the PEV can be expressed by $\alpha$, $\bar{Y}_c^{un}$ and $\pi_R$ as follows, drawing on theorem 1:

$$PEV = P(\delta < \delta^{\#} \mid \mathbf{D^{id}}) =$$

$$P\left( \frac{\delta - (\bar{Y}_t^{id} - \bar{Y}_c^{id})}{\sqrt{\frac{\sigma_t^2}{n_t^{ob} + n_t^{un}} + \frac{\sigma_c^2}{n_c^{ob} + n_c^{un}}}} < \frac{\delta^{\#} - (\bar{Y}_t^{id} - \bar{Y}_c^{id})}{\sqrt{\frac{\sigma_t^2}{n_t^{ob} + n_t^{un}} + \frac{\sigma_c^2}{n_c^{ob} + n_c^{un}}}} \mid \mathbf{D^{id}} \right) = \Phi\left( \frac{\delta^{\#} - (\bar{Y}_t^{id} - \bar{Y}_c^{id})}{\sqrt{\frac{\sigma_t^2}{n_t^{ob} + n_t^{un}} + \frac{\sigma_c^2}{n_c^{ob} + n_c^{un}}}} \right)$$

$$= \Phi\left( \frac{\delta^{\#} - [(1-\pi_R)\alpha \bar{Y}_c^{un} + \pi_R \bar{Y}_t^{ob} - (1-\pi_R)\bar{Y}_c^{un} - \pi_R \bar{Y}_c^{ob}]}{\sqrt{\pi_R \frac{\sigma_t^2}{n_t^{ob}} + \pi_R \frac{\sigma_c^2}{n_c^{ob}}}} \right)$$

$$= \Phi\left( \frac{\bar{Y}_c^{un} \cdot \alpha\pi_R - \bar{Y}_c^{un} \cdot \alpha - (\bar{Y}_t^{ob} - \bar{Y}_c^{ob} + \bar{Y}_c^{un}) \cdot \pi_R + (\bar{Y}_c^{un} + \delta^{\#})}{\sqrt{\pi_R \left( \frac{\sigma_t^2}{n_t^{ob}} + \frac{\sigma_c^2}{n_c^{ob}} \right)}} \right)$$

(A5)

From (A5), the probit model for PEV can be derived as identical to (6).

Likewise, when a significant negative effect has been concluded, the PEV is expressed as follows:

$$PEV = P(\delta > \delta^{\#} \mid \mathbf{D^{id}}) =$$

$$P\left( \frac{\delta - (\bar{Y}_t^{id} - \bar{Y}_c^{id})}{\sqrt{\frac{\sigma_t^2}{n_t^{ob} + n_t^{un}} + \frac{\sigma_c^2}{n_c^{ob} + n_c^{un}}}} > \frac{\delta^{\#} - (\bar{Y}_t^{id} - \bar{Y}_c^{id})}{\sqrt{\frac{\sigma_t^2}{n_t^{ob} + n_t^{un}} + \frac{\sigma_c^2}{n_c^{ob} + n_c^{un}}}} \mid \mathbf{D^{id}} \right) = 1 - \Phi\left( \frac{\delta^{\#} - (\bar{Y}_t^{id} - \bar{Y}_c^{id})}{\sqrt{\frac{\sigma_t^2}{n_t^{ob} + n_t^{un}} + \frac{\sigma_c^2}{n_c^{ob} + n_c^{un}}}} \right)$$

$$= \Phi\left( \frac{[(1-\pi_R)\alpha \bar{Y}_c^{un} + \pi_R \bar{Y}_t^{ob} - (1-\pi_R)\bar{Y}_c^{un} - \pi_R \bar{Y}_c^{ob}] - \delta^{\#}}{\sqrt{\pi_R \frac{\sigma_t^2}{n_t^{ob}} + \pi_R \frac{\sigma_c^2}{n_c^{ob}}}} \right)$$

$$= \Phi\left( \frac{\bar{Y}_c^{un} \cdot \alpha + (\bar{Y}_t^{ob} - \bar{Y}_c^{ob} + \bar{Y}_c^{un}) \cdot \pi_R - \bar{Y}_c^{un} \cdot \alpha\pi_R - (\bar{Y}_c^{un} + \delta^{\#})}{\sqrt{\pi_R \left( \frac{\sigma_t^2}{n_t^{ob}} + \frac{\sigma_c^2}{n_c^{ob}} \right)}} \right)$$

(A6)

From (A6), the probit model for PEV can be derived as identical to (7).

Proof of theorem 3:

Given the Bayesian model in (8), the posterior distribution for $\mu_t$ is as follows:

$$\mu_t \mid \mathbf{D^{ob}} \sim N\left(\frac{n_t}{N_t + n_t} \bar{Y}_t^{un} + \frac{N_t}{N_t + n_t} \bar{Y}_t^{ob}, \frac{\sigma_t^2}{N_t + n_t}\right) \tag{A7}$$

Furthermore, the posterior distribution for $\mu_c$ is as follows:

$$\mu_c \mid \mathbf{D^{ob}} \sim N\left(\frac{n_c}{N_c + n_c} \bar{Y}_c^{un} + \frac{N_c}{N_c + n_c} \bar{Y}_c^{ob}, \frac{\sigma_c^2}{N_c + n_c}\right) \tag{A8}$$

Assuming the treated and the control outcomes are independent, the posterior distribution of $\delta$ is equivalent to $\mu_t \mid \mathbf{D^{ob}} - \mu_c \mid \mathbf{D^{ob}}$ and therefore it has the identical form as (9).

Proof of theorem 4:

My goal is to derive the formula for least square estimate of regression coefficient for W (i.e., $\hat{\beta}_W$) based on the ideal sample and the variance of $\hat{\beta}_W$, since they are the mean and the variance of the distribution of $\delta$ conditional on the ideal sample. First, I need to define the following ordered data matrices for the ideal sample:

$$\begin{aligned}
\mathbf{D} &= [\mathbf{Y}_{(n^{un}+n^{ob})\times 1}, \mathbf{X}_{(n^{un}+n^{ob})\times(p+2)}] \\
\mathbf{X} &= [\mathbf{1}_{(n^{un}+n^{ob})\times 1}, \mathbf{V}_{(n^{un}+n^{ob})\times(p+1)}] \\
\mathbf{V} &= [\mathbf{Z}_{(n^{un}+n^{ob})\times p}, \mathbf{W}_{(n^{un}+n^{ob})\times 1}] \\
\mathbf{Z} &= [\mathbf{Z_1}, \mathbf{Z_2}, ..., \mathbf{Z_p}]_{(n^{un}+n^{ob})\times p}
\end{aligned} \tag{A9}$$

and the following ordered mean vectors:

$$\begin{aligned}
\bar{\mathbf{V}}^{\mathbf{id}} &= [\bar{\mathbf{Z}}^{\mathbf{id}}, \bar{W}^{id}]_{1\times(p+1)} \\
\bar{\mathbf{Z}}^{\mathbf{id}} &= [\bar{Z}_1^{id}, \bar{Z}_2^{id}, \cdots, \bar{Z}_p^{id}]_{1\times p}
\end{aligned} \tag{A10}$$

The matrix $\mathbf{X^T X}$ for the ideal sample could then be molded as the following block matrix:

$$\mathbf{X^T X} = \begin{pmatrix} n^{un}+n^{ob} & (n^{un}+n^{ob})\bar{\mathbf{V}}^{\mathbf{id}} \\ (n^{un}+n^{ob})(\bar{\mathbf{V}}^{\mathbf{id}})^{\mathbf{T}} & \mathbf{V^T V} \end{pmatrix} \tag{A11}$$

The inverse of $\mathbf{X^T X}$ can be shown to have the following form:

$$(\mathbf{X}^{\mathbf{T}}\mathbf{X})^{\mathbf{-1}} = \begin{pmatrix} \frac{1}{n^{un}+n^{ob}} + \bar{\mathbf{V}}^{\mathbf{id}} \frac{1}{n^{un}+n^{ob}} (\mathbf{S}^{\mathbf{id}}_{\mathbf{VV}})^{\mathbf{-1}} (\bar{\mathbf{V}}^{\mathbf{id}})^{\mathbf{T}} & -\frac{1}{n^{un}+n^{ob}} \bar{\mathbf{V}}^{\mathbf{id}} (\mathbf{S}^{\mathbf{id}}_{\mathbf{VV}})^{\mathbf{-1}} \\ -\frac{1}{n^{un}+n^{ob}} (\mathbf{S}^{\mathbf{id}}_{\mathbf{VV}})^{\mathbf{-1}} (\bar{\mathbf{V}}^{\mathbf{id}})^{\mathbf{T}} & \frac{1}{n^{un}+n^{ob}} (\mathbf{S}^{\mathbf{id}}_{\mathbf{VV}})^{\mathbf{-1}} \end{pmatrix}$$

(A12)

It should be clear now that, to determine the definite form of $(\mathbf{X}^{\mathbf{T}}\mathbf{X})^{\mathbf{-1}}$ I need to find out what $(\mathbf{S}^{\mathbf{id}}_{\mathbf{VV}})^{\mathbf{-1}}$ is. As a variance-covariance matrix for the vector of predictors V, $\mathbf{S}^{\mathbf{id}}_{\mathbf{VV}}$ can be expressed as the block matrix whose elements is formalized below:

$$\mathbf{S}^{\mathbf{id}}_{\mathbf{VV}} = \begin{pmatrix} \mathbf{S}^{\mathbf{id}}_{\mathbf{ZZ}} & \mathbf{S}^{\mathbf{id}}_{\mathbf{ZW}} \\ \mathbf{S}^{\mathbf{id}}_{\mathbf{WZ}} & \hat{\sigma}^{id}_{WW} \end{pmatrix}_{(p+1)\times(p+1)} \tag{A13}$$

where:

$$\mathbf{S}^{\mathbf{id}}_{\mathbf{ZZ}} = \begin{pmatrix} \hat{\sigma}^{id}_{Z_1 Z_1} & \cdots & \hat{\sigma}^{id}_{Z_1 Z_p} \\ \vdots & \ddots & \vdots \\ \hat{\sigma}^{id}_{Z_p Z_1} & \cdots & \hat{\sigma}^{id}_{Z_p Z_p} \end{pmatrix}_{p\times p}$$

$$\mathbf{S}^{\mathbf{id}}_{\mathbf{ZW}} = \begin{pmatrix} \hat{\sigma}^{id}_{Z_1 W} \\ \vdots \\ \hat{\sigma}^{id}_{Z_p W} \end{pmatrix}_{p\times 1} \tag{A14}$$

$$\mathbf{S}^{\mathbf{id}}_{\mathbf{WZ}} = \begin{pmatrix} \hat{\sigma}^{id}_{Z_1 W} & \cdots & \hat{\sigma}^{id}_{Z_p W} \end{pmatrix}_{1\times p}$$

Furthermore, I define the following covariance vector:

$$\mathbf{S}_{\mathbf{ZY}}^{\mathbf{id}} = \begin{pmatrix} \hat{\sigma}_{Z_1 Y}^{id} \\ \vdots \\ \hat{\sigma}_{Z_p Y}^{id} \end{pmatrix}_{p\times 1} \tag{A15}$$

All aforementioned sample covariances and variances are supposed to be computed according to the following formula:

$$\begin{aligned} \hat{\sigma}_{xy} &= \frac{1}{n}\sum_{i=1}^{n}(x_i - \overline{x})(y_i - \overline{y}) \\ \hat{\sigma}_{xx} &= \frac{1}{n}\sum_{i=1}^{n}(x_i - \overline{x})^2 \end{aligned} \tag{A16}$$

for any variable x or y and any sample size n in this context.

Consequently, the inverse of $\mathbf{S}_{\mathbf{VV}}^{\mathbf{id}}$ can be formulated here:

$$(\mathbf{S}_{\mathbf{VV}}^{\mathbf{id}})^{-1} =$$

$$\begin{bmatrix} (\mathbf{S}_{\mathbf{ZZ}}^{\mathbf{id}})^{-1} + (\mathbf{S}_{\mathbf{ZZ}}^{\mathbf{id}})^{-1}\mathbf{S}_{\mathbf{ZW}}^{\mathbf{id}}(\hat{\sigma}_{WW}^{id} - \mathbf{S}_{\mathbf{WZ}}^{\mathbf{id}}(\mathbf{S}_{\mathbf{ZZ}}^{\mathbf{id}})^{-1}\mathbf{S}_{\mathbf{ZW}}^{\mathbf{id}})^{-1}\mathbf{S}_{\mathbf{WZ}}^{\mathbf{id}}(\mathbf{S}_{\mathbf{ZZ}}^{\mathbf{id}})^{-1} & -(\mathbf{S}_{\mathbf{ZZ}}^{\mathbf{id}})^{-1}\mathbf{S}_{\mathbf{ZW}}^{\mathbf{id}}(\hat{\sigma}_{WW}^{id} - \mathbf{S}_{\mathbf{WZ}}^{\mathbf{id}}(\mathbf{S}_{\mathbf{ZZ}}^{\mathbf{id}})^{-1}\mathbf{S}_{\mathbf{ZW}}^{\mathbf{id}})^{-1} \\ -(\hat{\sigma}_{WW}^{id} - \mathbf{S}_{\mathbf{WZ}}^{\mathbf{id}}(\mathbf{S}_{\mathbf{ZZ}}^{\mathbf{id}})^{-1}\mathbf{S}_{\mathbf{ZW}}^{\mathbf{id}})^{-1}\mathbf{S}_{\mathbf{WZ}}^{\mathbf{id}}(\mathbf{S}_{\mathbf{ZZ}}^{\mathbf{id}})^{-1} & (\hat{\sigma}_{WW}^{id} - \mathbf{S}_{\mathbf{WZ}}^{\mathbf{id}}(\mathbf{S}_{\mathbf{ZZ}}^{\mathbf{id}})^{-1}\mathbf{S}_{\mathbf{ZW}}^{\mathbf{id}})^{-1} \end{bmatrix}$$

(A17)

Plugging the above matrix of $(\mathbf{S}_{\mathbf{VV}}^{\mathbf{id}})^{-1}$ into the block matrix of $(\mathbf{X}^{\mathbf{T}}\mathbf{X})^{-1}$ will give us the complete definite form of matrix $(\mathbf{X}^{\mathbf{T}}\mathbf{X})^{-1}$, whose elements are all ideal sample statistics such as ideal sample variances, ideal sample covariances and ideal sample means. To isolate the estimated regression coefficient for W, I only need to use the elements in the last row of $(\mathbf{X}^{\mathbf{T}}\mathbf{X})^{-1}$, which are provide next:

$$(\mathbf{X^TX})^{\mathbf{-1}}_{(p+2)1} = \frac{1}{n^{un}+n^{ob}}[(\hat{\sigma}^{id}_{WW} - \mathbf{S^{id}_{WZ}}(\mathbf{S^{id}_{ZZ}})^{-1}\mathbf{S^{id}_{ZW}})^{-1}\mathbf{S^{id}_{WZ}}(\mathbf{S^{id}_{ZZ}})^{-1}(\bar{\mathbf{Z}}^{\mathbf{id}})^{\mathbf{T}} - \bar{W}^{id}(\hat{\sigma}^{id}_{WW} - \mathbf{S^{id}_{WZ}}(\mathbf{S^{id}_{ZZ}})^{-1}\mathbf{S^{id}_{ZW}})^{-1}]$$

$$[(\mathbf{X^TX})^{\mathbf{-1}}_{(p+2)2},\cdots,(\mathbf{X^TX})^{\mathbf{-1}}_{(p+2)(p+1)}] = -\frac{1}{n^{un}+n^{ob}}(\hat{\sigma}^{id}_{WW} - \mathbf{S^{id}_{WZ}}(\mathbf{S^{id}_{ZZ}})^{-1}\mathbf{S^{id}_{ZW}})^{-1}\mathbf{S^{id}_{WZ}}(\mathbf{S^{id}_{ZZ}})^{-1}$$

$$(\mathbf{X^TX})^{\mathbf{-1}}_{(p+2)(p+2)} = \frac{1}{n^{un}+n^{ob}}(\hat{\sigma}^{id}_{WW} - \mathbf{S^{id}_{WZ}}(\mathbf{S^{id}_{ZZ}})^{-1}\mathbf{S^{id}_{ZW}})^{-1}$$

(A18)

Because the estimated regression coefficient for W is the last element of $(\mathbf{X^TX})^{\mathbf{-1}}\mathbf{X^TY}$ which is the dot product between the last row of $(\mathbf{X^TX})^{\mathbf{-1}}$ and $\mathbf{X^TY}$, the expression of $\mathbf{X^TY}$ is also needed here:

$$\mathbf{X^TY} = \begin{bmatrix} (n^{un}+n^{ob})\bar{Y}^{id} \\ \mathbf{Z^TY} \\ \mathbf{W^TY} \end{bmatrix} \tag{A19}$$

where:

$$\begin{aligned} \mathbf{Z^TY} &= (n^{un}+n^{ob})\mathbf{S^{id}_{ZY}} + (n^{un}+n^{ob})\bar{Y}^{id}(\bar{\mathbf{Z}}^{\mathbf{id}})^{\mathbf{T}} \\ \mathbf{W^TY} &= (n^{un}+n^{ob})\hat{\sigma}^{id}_{WY} + (n^{un}+n^{ob})\bar{W}^{id}\bar{Y}^{id} \end{aligned} \tag{A20}$$

Now I can calculate the estimated regression coefficient for W as the dot product between the last row of $(\mathbf{X^TX})^{\mathbf{-1}}$ and the vector $\mathbf{X^TY}$. The result is presented below:

$$\hat{\beta}_W = \frac{\hat{\sigma}^{id}_{WY} - \mathbf{S^{id}_{WZ}}(\mathbf{S^{id}_{ZZ}})^{-1}\mathbf{S^{id}_{ZY}}}{\hat{\sigma}^{id}_{WW} - \mathbf{S^{id}_{WZ}}(\mathbf{S^{id}_{ZZ}})^{-1}\mathbf{S^{id}_{ZW}}} \tag{A21}$$

The variance of $\hat{\beta}_W$ should be straightforward: it is just the product of the known residual variance $\sigma^2$ and the element in the (p+2)th row and the (p+2)th column of $(\mathbf{X}^{\mathbf{T}}\mathbf{X})^{-1}$ :

$$Var(\hat{\beta}_W) = \frac{\sigma^2}{n^{un}+n^{ob}} (\hat{\sigma}_{WW}^{id} - \mathbf{S_{WZ}^{id}}(\mathbf{S_{ZZ}^{id}})^{-1}\mathbf{S_{ZW}^{id}})^{-1} \tag{A22}$$

Taken together, the distribution of $\delta$ given the ideal sample for the simple estimator is:

$$\delta \,|\, \mathbf{D^{id}} \sim N\left(\frac{\hat{\sigma}_{WY}^{id} - \mathbf{S_{WZ}^{id}}(\mathbf{S_{ZZ}^{id}})^{-1}\mathbf{S_{ZY}^{id}}}{\hat{\sigma}_{WW}^{id} - \mathbf{S_{WZ}^{id}}(\mathbf{S_{ZZ}^{id}})^{-1}\mathbf{S_{ZW}^{id}}}, \frac{\sigma^2}{n^{un}+n^{ob}} (\hat{\sigma}_{WW}^{id} - \mathbf{S_{WZ}^{id}}(\mathbf{S_{ZZ}^{id}})^{-1}\mathbf{S_{ZW}^{id}})^{-1}\right)$$

(A23)

The derivations of ideal variances/covariances as functions of observed and unobserved sample statistics follow the same reasoning. Here I just take the covariance between W and Y in the ideal sample as an example. First of all, I have:

$$\hat{\sigma}_{WY}^{id} = \frac{1}{n^{un}+n^{ob}} \sum_{i=1}^{n^{un}+n^{ob}} (W_i - \bar{W}^{id})(Y_i - \bar{Y}^{id}) \tag{A24}$$

(A24) can be rearranged and re-expressed as follows:

$$\sum_{i=1}^{n^{un}+n^{ob}} W_i Y_i = \left(n^{un}+n^{ob}\right)\hat{\sigma}_{WY}^{id} + \left(n^{un}+n^{ob}\right)\bar{W}^{id}\bar{Y}^{id} = \sum_{i=1}^{n^{un}} W_i Y_i + \sum_{i=1}^{n^{ob}} W_i Y_i$$

(A25)

Similarly, the following equations are true for the observed sample and the unobserved sample:

$$\sum_{i=1}^{n^{un}} W_i Y_i = n^{un}\hat{\sigma}_{WY}^{un} + n^{un}\bar{W}^{un}\bar{Y}^{un}$$

$$\sum_{i=1}^{n^{ob}} W_i Y_i = n^{ob}\hat{\sigma}_{WY}^{ob} + n^{ob}\bar{W}^{ob}\bar{Y}^{ob} \qquad \text{(A26)}$$

(A25) and (A26) could be consolidated into an expanded one as follows:

$$\left(n^{un}+n^{ob}\right)\hat{\sigma}_{WY}^{id} + \left(n^{un}+n^{ob}\right)\bar{W}^{id}\bar{Y}^{id} = n^{un}\hat{\sigma}_{WY}^{un} + n^{ob}\hat{\sigma}_{WY}^{ob} + n^{un}\bar{W}^{un}\bar{Y}^{un} + n^{ob}\bar{W}^{ob}\bar{Y}^{ob}$$

(A27)

Finally, the expression of $\hat{\sigma}_{WY}^{id}$ as a function of unobserved and observed sample statistics could be deduced from (A27):

$$\begin{aligned}
\hat{\sigma}_{WY}^{id} &= \frac{n^{un}\hat{\sigma}_{WY}^{un} + n^{ob}\hat{\sigma}_{WY}^{ob}}{n^{un}+n^{ob}} + \frac{n^{un}\bar{W}^{un}\bar{Y}^{un} + n^{ob}\bar{W}^{ob}\bar{Y}^{ob}}{n^{un}+n^{ob}} - \bar{W}^{id}\bar{Y}^{id} \\
&= \frac{n^{un}}{n^{un}+n^{ob}}\hat{\sigma}_{WY}^{un} + \frac{n^{ob}}{n^{un}+n^{ob}}\hat{\sigma}_{WY}^{ob} + \frac{n^{un}}{n^{un}+n^{ob}}\bar{W}^{un}\bar{Y}^{un} + \frac{n^{ob}}{n^{un}+n^{ob}}\bar{W}^{ob}\bar{Y}^{ob} \\
&- \left(\frac{n^{un}\bar{W}^{un} + n^{ob}\bar{W}^{ob}}{n^{un}+n^{ob}}\right)\left(\frac{n^{un}\bar{Y}^{un} + n^{ob}\bar{Y}^{ob}}{n^{un}+n^{ob}}\right) \\
&= \lambda\hat{\sigma}_{WY}^{un} + (1-\lambda)\hat{\sigma}_{WY}^{ob} + \lambda\bar{W}^{un}\bar{Y}^{un} + (1-\lambda)\bar{W}^{ob}\bar{Y}^{ob} \\
&- \left[\lambda^2\bar{W}^{un}\bar{Y}^{un} + \lambda(1-\lambda)\bar{W}^{un}\bar{Y}^{ob} + \lambda(1-\lambda)\bar{W}^{ob}\bar{Y}^{un} + (1-\lambda)^2\bar{W}^{ob}\bar{Y}^{ob}\right] \\
&= \lambda\hat{\sigma}_{WY}^{un} + (1-\lambda)\hat{\sigma}_{WY}^{ob} + \lambda(1-\lambda)[\bar{W}^{un}\bar{Y}^{un} - \bar{W}^{un}\bar{Y}^{ob} - \bar{W}^{ob}\bar{Y}^{un} + \bar{W}^{ob}\bar{Y}^{ob}] \\
&= \lambda\hat{\sigma}_{WY}^{un} + (1-\lambda)\hat{\sigma}_{WY}^{ob} + (1-\lambda)\lambda(\bar{W}^{ob} - \bar{W}^{un})(\bar{Y}^{ob} - \bar{Y}^{un})
\end{aligned} \qquad \text{(A28)}$$

where:

$$\lambda = \frac{n^{un}}{n^{un}+n^{ob}} \qquad \text{(A29)}$$

Proof of theorem 5:

Based on theorem 4, the PEV is shown to be a function of the ideal sample statistics. Specifically, when a significant positive effect has been concluded, the PEV is written as follows:

$$PEV = P(\delta < \delta^{\#} \mid \mathbf{D^{id}}) =$$

$$P\left( \frac{\delta - \dfrac{\hat{\sigma}_{WY}^{id} - \mathbf{S_{WZ}^{id}}(\mathbf{S_{ZZ}^{id}})^{-1}\mathbf{S_{ZY}^{id}}}{\hat{\sigma}_{WW}^{id} - \mathbf{S_{WZ}^{id}}(\mathbf{S_{ZZ}^{id}})^{-1}\mathbf{S_{ZW}^{id}}}}{\sqrt{\dfrac{\sigma^2}{n^{un}+n^{ob}}(\hat{\sigma}_{WW}^{id} - \mathbf{S_{WZ}^{id}}(\mathbf{S_{ZZ}^{id}})^{-1}\mathbf{S_{ZW}^{id}})^{-1}}} < \frac{\delta^{\#} - \dfrac{\hat{\sigma}_{WY}^{id} - \mathbf{S_{WZ}^{id}}(\mathbf{S_{ZZ}^{id}})^{-1}\mathbf{S_{ZY}^{id}}}{\hat{\sigma}_{WW}^{id} - \mathbf{S_{WZ}^{id}}(\mathbf{S_{ZZ}^{id}})^{-1}\mathbf{S_{ZW}^{id}}}}{\sqrt{\dfrac{\sigma^2}{n^{un}+n^{ob}}(\hat{\sigma}_{WW}^{id} - \mathbf{S_{WZ}^{id}}(\mathbf{S_{ZZ}^{id}})^{-1}\mathbf{S_{ZW}^{id}})^{-1}}} \mid \mathbf{D^{id}} \right)$$

$$= \Phi\left( \frac{\delta^{\#} - \dfrac{\hat{\sigma}_{WY}^{id} - \mathbf{S_{WZ}^{id}}(\mathbf{S_{ZZ}^{id}})^{-1}\mathbf{S_{ZY}^{id}}}{\hat{\sigma}_{WW}^{id} - \mathbf{S_{WZ}^{id}}(\mathbf{S_{ZZ}^{id}})^{-1}\mathbf{S_{ZW}^{id}}}}{\sqrt{\dfrac{\sigma^2}{n^{un}+n^{ob}}(\hat{\sigma}_{WW}^{id} - \mathbf{S_{WZ}^{id}}(\mathbf{S_{ZZ}^{id}})^{-1}\mathbf{S_{ZW}^{id}})^{-1}}} \right)$$

$$= \Phi\left( \frac{\sqrt{n^{un}+n^{ob}}}{\sigma\sqrt{\hat{\sigma}_{WW}^{id} - \mathbf{S_{WZ}^{id}}(\mathbf{S_{ZZ}^{id}})^{-1}\mathbf{S_{ZW}^{id}}}} [\delta^{\#}(\hat{\sigma}_{WW}^{id} - \mathbf{S_{WZ}^{id}}(\mathbf{S_{ZZ}^{id}})^{-1}\mathbf{S_{ZW}^{id}}) - (\hat{\sigma}_{WY}^{id} - \mathbf{S_{WZ}^{id}}(\mathbf{S_{ZZ}^{id}})^{-1}\mathbf{S_{ZY}^{id}})] \right)$$

(A30)

From (A30), the probit model for the PEV can be shown to be identical to (15).

Likewise, when a significant negative effect has been concluded, the PEV is expressed as follows:

$$PEV = P(\delta > \delta^{\#} \mid \mathbf{D^{id}}) =$$

$$P\left( \frac{\delta - \dfrac{\hat{\sigma}_{WY}^{id} - \mathbf{S_{WZ}^{id}}(\mathbf{S_{ZZ}^{id}})^{-1}\mathbf{S_{ZY}^{id}}}{\hat{\sigma}_{WW}^{id} - \mathbf{S_{WZ}^{id}}(\mathbf{S_{ZZ}^{id}})^{-1}\mathbf{S_{ZW}^{id}}}}{\sqrt{\dfrac{\sigma^2}{n^{un}+n^{ob}}(\hat{\sigma}_{WW}^{id} - \mathbf{S_{WZ}^{id}}(\mathbf{S_{ZZ}^{id}})^{-1}\mathbf{S_{ZW}^{id}})^{-1}}} > \frac{\delta^{\#} - \dfrac{\hat{\sigma}_{WY}^{id} - \mathbf{S_{WZ}^{id}}(\mathbf{S_{ZZ}^{id}})^{-1}\mathbf{S_{ZY}^{id}}}{\hat{\sigma}_{WW}^{id} - \mathbf{S_{WZ}^{id}}(\mathbf{S_{ZZ}^{id}})^{-1}\mathbf{S_{ZW}^{id}}}}{\sqrt{\dfrac{\sigma^2}{n^{un}+n^{ob}}(\hat{\sigma}_{WW}^{id} - \mathbf{S_{WZ}^{id}}(\mathbf{S_{ZZ}^{id}})^{-1}\mathbf{S_{ZW}^{id}})^{-1}}} \mid \mathbf{D^{id}} \right)$$

$$= 1 - \Phi\left( \frac{\delta^{\#} - \dfrac{\hat{\sigma}_{WY}^{id} - \mathbf{S_{WZ}^{id}}(\mathbf{S_{ZZ}^{id}})^{-1}\mathbf{S_{ZY}^{id}}}{\hat{\sigma}_{WW}^{id} - \mathbf{S_{WZ}^{id}}(\mathbf{S_{ZZ}^{id}})^{-1}\mathbf{S_{ZW}^{id}}}}{\sqrt{\dfrac{\sigma^2}{n^{un}+n^{ob}}(\hat{\sigma}_{WW}^{id} - \mathbf{S_{WZ}^{id}}(\mathbf{S_{ZZ}^{id}})^{-1}\mathbf{S_{ZW}^{id}})^{-1}}} \right)$$

$$= \Phi\left( \frac{\sqrt{n^{un}+n^{ob}}}{\sigma\sqrt{\hat{\sigma}_{WW}^{id} - \mathbf{S_{WZ}^{id}}(\mathbf{S_{ZZ}^{id}})^{-1}\mathbf{S_{ZW}^{id}}}}[(\hat{\sigma}_{WY}^{id} - \mathbf{S_{WZ}^{id}}(\mathbf{S_{ZZ}^{id}})^{-1}\mathbf{S_{ZY}^{id}}) - \delta^{\#}(\hat{\sigma}_{WW}^{id} - \mathbf{S_{WZ}^{id}}(\mathbf{S_{ZZ}^{id}})^{-1}\mathbf{S_{ZW}^{id}})] \right)$$

(A31)

From (A31), the probit model for the PEV can be shown to be identical to (16).

Proof of theorem 6:

Given the Bayesian model in (17), the posterior distribution of $\boldsymbol{\beta}$ is as follows:

$$\beta \mid \mathbf{D^{ob}} \sim N(\boldsymbol{\theta}_{\boldsymbol{\beta}}, \boldsymbol{\Phi}_{\boldsymbol{\beta}}) \qquad \text{(A32)}$$

where:

$$\begin{aligned} \boldsymbol{\theta}_{\boldsymbol{\beta}} &= ((\mathbf{X^{un}})^{\mathbf{T}}\mathbf{X^{un}} + (\mathbf{X^{ob}})^{\mathbf{T}}\mathbf{X^{ob}})^{-1}((\mathbf{X^{un}})^{\mathbf{T}}\mathbf{Y^{un}} + (\mathbf{X^{ob}})^{\mathbf{T}}\mathbf{Y^{ob}}) \\ \boldsymbol{\Phi}_{\boldsymbol{\beta}} &= \sigma^2((\mathbf{X^{un}})^{\mathbf{T}}\mathbf{X^{un}} + (\mathbf{X^{ob}})^{\mathbf{T}}\mathbf{X^{ob}})^{-1} \end{aligned} \qquad \text{(A33)}$$

Moreover, the following equations hold for the ideal sample:

$$\begin{aligned}(\mathbf{X}^{\mathbf{id}})^{\mathbf{T}}\mathbf{X}^{\mathbf{id}} &= (\mathbf{X}^{\mathbf{un}})^{\mathbf{T}}\mathbf{X}^{\mathbf{un}} + (\mathbf{X}^{\mathbf{ob}})^{\mathbf{T}}\mathbf{X}^{\mathbf{ob}} \\ (\mathbf{X}^{\mathbf{id}})^{\mathbf{T}}\mathbf{Y}^{\mathbf{id}} &= (\mathbf{X}^{\mathbf{un}})^{\mathbf{T}}\mathbf{Y}^{\mathbf{un}} + (\mathbf{X}^{\mathbf{ob}})^{\mathbf{T}}\mathbf{Y}^{\mathbf{ob}}\end{aligned} \tag{A34}$$

Combining (A33) and (A34), it is straightforward that the posterior distribution of $\boldsymbol{\beta}$ is identical to the distribution of $\boldsymbol{\beta}$ given the ideal sample, which has the mean as $((\mathbf{X}^{\mathbf{id}})^{\mathbf{T}}\mathbf{X}^{\mathbf{id}})^{-1}((\mathbf{X}^{\mathbf{id}})^{\mathbf{T}}\mathbf{Y}^{\mathbf{id}})$ and the variance as $\sigma^2((\mathbf{X}^{\mathbf{id}})^{\mathbf{T}}\mathbf{X}^{\mathbf{id}})^{-1}$. The posterior distribution of $\delta$, which is just the marginal distribution of $\beta_W$ in (A32), should then be identical to (13) which is the marginal distribution of $\beta_W$ in the distribution of $\boldsymbol{\beta}$ given the ideal sample.